\documentclass[aip,reprint,onecolumn]{revtex4-1}
\usepackage{longtable}
\usepackage{appendix}
\usepackage{amsmath}
\usepackage{graphicx}
\usepackage{lineno}
\usepackage{array}
\usepackage{longtable}
\usepackage{natbib}
\usepackage{subfigure}
\usepackage{bm}

\draft 

\begin{document}


\title{Penetrative magneto-convection of a rotating Boussinesq flow in $f$-planes} 




\author{Fan Xu}
\affiliation{State Key Laboratory of Lunar and Planetary Sciences, Macau University of Science and Technology, Macau, People's Republic of China}

\author{Tao Cai}
\email{tcai@must.edu.mo; astroct@gmail.com}
\affiliation{State Key Laboratory of Lunar and Planetary Sciences, Macau University of Science and Technology, Macau, People's Republic of China}


\date{\today}

\begin{abstract}
In this study, we conducted a linear instability analysis of penetrative magneto-convection in rapidly rotating Boussinesq flows within tilted f-planes, under the influence of a uniform background magnetic field. We integrated wave theory and convection theory to elucidate the penetration dynamics in rotating magneto-convection. Our findings suggest that efficient penetration in rapidly rotating flows with weakly stratified stable layers at low latitudes can be attributed to the resonance of wave transmission near the interface between unstable and stable layers. In the context of strongly stratified flows, we derived the scaling relationships of penetrative distances $\Delta$ with the stability parameter $\delta$. Our calculation shows that, for both rotation-dominated and magnetism-dominated flows, $\Delta$ obeys a scaling of $\Delta\sim O(\delta^{-1/2})$. In rotation-dominated flows, we noted a general decrease in penetrative distance with increased rotational effect, and a minor decrease in penetrative distance with increased latitude. When a background magnetic field is introduced, we observed a significant shift in penetrative distance as the Elsasser number $\Lambda$ approaches one. The penetrative distance tends to decrease when $\Lambda \ll 1$ and increase when $\Lambda \gg 1$ with the rotational effect, indicating a transition from rotation-dominated to magnetism-dominated flow. We have further investigated the impact of the background magnetic field when it is not aligned with the rotational axis. This presents a notable contrast to the case where the magnetic field is parallel to the rotational axis.
\end{abstract}


\maketitle 

Thermal convection occurs in many stars and planets, involving the penetration of convective flow from a convectively unstable layer to its adjacent stable layer. Penetrative convection has been widely studied due to its geophysical and astrophysical applications. However, estimating the penetrative distance remains a longstanding question. Previous investigations have mainly focused on penetrative convection without considering the rotational and magnetic effects. For example, theoretical analyses on penetrative distance in stellar convection have been performed based on mixing length theory \citep{zahn1991convective}, and Reynolds stress models \citep{xiong1997nonlocal,canuto2011stellar,zhang2012turbulent}. The penetrative structures can be different as predicted by different models \citep{zhang2012turbulent}. Numerical simulations \citep{brummell2002penetration,K2019Overshooting,korre2019convective,cai2020aupward,cai2020bupward,anders2022stellar} have also been extensively employed to study penetrative convection in stars and planets, complementing theoretical analyses. Some simulation results \citep{anders2022stellar} support the theory proposed by \citet{zahn1991convective}, while there are other simulations \citep{cai2020bupward,cai2020cupward} support the theoretical model proposed by \citet{zhang2012turbulent}. Most simulations employ idealized models to establish scaling relations between the penetrative distance and other physical variables. These scalings are then extrapolated to real stars \citep{hurlburt1994penetration,freytag1996hydrodynamical}. Simulations on penetrative convection are challenging when realistic values are used. Some examples are shown in \citet{K2019Overshooting} by using self-consistent opacity table, \citet{hotta2017solar} by using realistic energy flux. A few attempts have also been made to simulate penetrative convection by using realistic stellar structures \citep{arnett2015beyond,baraffe2023study}. Results indicate that the scalings obtained by idealized model may deviate from those in real stars \citep{baraffe2023study}. Consequently, it is advisable to utilize parameters that closely mirror reality.

While the majority of stars and planets exhibit rotation and many possess magnetic fields, the study of penetrative magneto-convection within a rotating flow is still in its early stages. Rotational and magnetic effects are typically three-dimensional, posing challenges to theoretical modelling on their effects to penetrative convection. \citet{augustson2019model} proposed an analytical model for rotating penetration, specifically focusing on the dependency of the penetrative depth on the convective Rossby number and diffusivity of the flow. The model predicts a reduction in penetrative distance as the rotational effect intensifies. This is consistent with a recent simulation by \citet{fuentes2023rotation}, which also demonstrates a tendency for rotation to diminish penetrative distance. Simulation of penetration in rotating tilted $f$-plane have been performed by \citet{Pal2007Turbulent} and \citet{Pal2008Turbulent}, who noted that the influence of rotational effect on penetrative distance differ significantly between downward and upward penetration. In line with the theoretical model, rotation tends to negatively affect the penetration in the downward penetration \citep{Pal2007Turbulent}. However, in case of upward penetration, the penetrative distance exhibits an increasing trend with rotation \citep{Pal2008Turbulent}. \citet{Pal2007Turbulent} and \citet{Pal2008Turbulent} have also shown that the angle between the angular velocity and gravity can induce variations in the penetrative distances. \citet{PJ2004Local} conducted extensive numerical experiments on rotating penetration within a tilted f-plane. Their findings revealed that the relationship between penetration and rotation can invert when moving from high to low latitudes.
Penetrative convection in rotating spherical geometry has been investigated by \citet{browning2004simulations}, \citet{augustson2012convection}, and \citet{brun2017differential}. Given that differential rotation can be driven in spherical geometry, the relationship between penetrative distance and rotation becomes complex and varies across different latitudes. \citet{ziegler2003box} and \citet{PJ2004Local} have investigated the penetration in rotating magneto-convection in Cartesian geometry. Their results suggest a consistent decrease in penetrative distance with an increase in the strength of the background magnetic field.

Beyond numerical simulations, linear stability analysis can offer valuable insights into penetrative convection. \citet{zhang2000teleconvection} and \citet{zhang2002from} conducted comprehensive global linear stability analyses on the penetration of rapidly rotating convection. Their research demonstrated that penetration in rapidly rotating convection can be highly effective when the stable layer is weakly stratified. Intriguingly, they found that under certain conditions, the flow motions in the stable layer can be more vigorous than in the unstable layer. This phenomenon has been termed as `teleconvection' \citep{zhang2000teleconvection}. \citet{cai2020penetrative} has confirmed that teleconvection can be observed in the tilted $f$-plane at low to mid latitudes. This underscores the significance of the nontraditional term of the Coriolis force in driving teleconvection. \citet{takehiro2001penetration} and \citet{takehiro2015penetration} utilized wave theory to elucidate the penetration observed in rotating convection in spherical geometry. They proposed that the penetrative distance could be approximated by the decay distance of the evanescent wave within the stable layer. \citet{takehiro2001penetration} and \citet{takehiro2015penetration} primarily focused on the penetration of rotating convection within a strongly stratified stable layer. In this context, the propagation of gravity waves in the stable layer is suppressed, considering the frequency of inertial waves in the unstable layer. With rapid rotation, inertial waves can propagate within the unstable layer. However, if the stable layer is not strongly stratified, gravito-inertial waves are permitted to propagate. In such scenarios, wave transmission occurs at the interface between the unstable and stable layers. \citet{wei2020wave} and \citet{cai2021Inertial} have demonstrated that wave transmission can be highly efficient when the stable layer is weakly stratified or when the tilted plane is at a critical latitude. Consequently, efficient penetration can be achieved through wave transmission. \citet{garai2022convective} found that travelling wave instability occurs at the onset of convection in rapid rotation with lateral temperature gradient. \citet{mukherjee2023thermal} revealed that stable layers have significantly effect on the topology and symmetry of generated magnetic field. \citet{sahoo2023onset} studied the onset of oscillatory magneto-convection with rapid rotation convection and spatially varying magnetic field. In this paper, we integrate both wave and convection theories to elucidate the penetration observed in rotating magneto-convection by linear stability analysis. We try to link the penetrative convection with wave transmission across the interface between convectively unstable and stable layers, as shown in \citet{cai2021Inertial} and \citet{cai2021enhancement}.

\section{The model}
\begin{figure}
	\includegraphics[width=0.6\columnwidth]{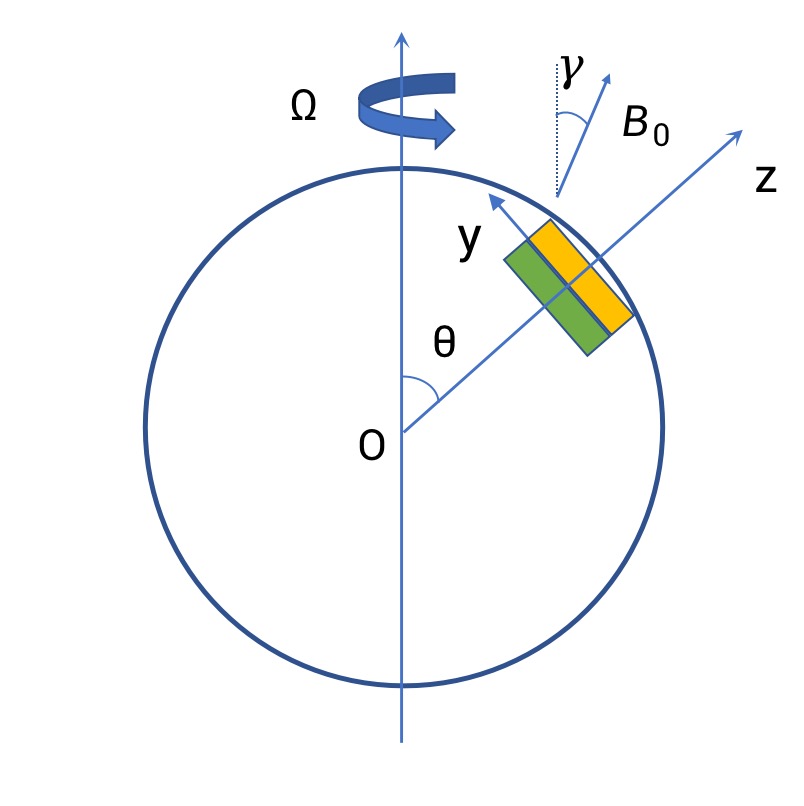}
    \caption{The sketch plot of the $f$-plane structure. It is adapted with permission from the reference Cai, The Astrophysical Journal, 898, 22 (2020) \citep{cai2020penetrative}. Copyright 2020 American Astronomical Society. The plane is inclined with the rotational axis at an angle $\theta$. The upper layer (yellow) is stable while the lower layer (green) is unstable. $\bm{\hat{z}}$ is the radial direction, $\bm{\hat{y}}$ is the south-north direction, $\bm{\hat{x}}$ is the west-east direction (pointing into the plane of the paper), $\bm{B}_{0}$ is the imposed background mean magnetic field. The angle between $\bm{B}_{0}$ and the rotational axis is $\gamma$.}
    \label{fig:f1}
\end{figure}

We consider a rotating Boussinesq flow with a mean background magnetic field in a tilted {\it f}-plane of colatitude $\theta$. In such a case, the governing equations for mass, momentum, energy conservation and magnetic induction can be written as
\begin{align}
    &\bm{\nabla} \bm{\cdot} \bm{u}=\bm{\nabla} \bm{\cdot} \bm{B}=0~,\\
    &\partial_{t} \bm{u}=-\bm{u}\cdot \nabla \bm{u} - 2\Omega  \bm{\hat{e}}_{\Omega}\bm{\times} \bm{u}-\rho^{-1}\nabla P+g\alpha T \bm{\hat{z}} + \nu \nabla^{2}\bm{u} ~\nonumber\\
    &\qquad+{(\mu\rho)}^{-1}(\nabla\times \bm{B})\times \bm{B}~,\\
    &\partial_{t} \bm{B}=\bm{\nabla} \bm{\times} (\bm{u} \bm{\times} \bm{B})+\eta \nabla^{2}\bm{B}~,\\
    &\partial_{t} T=-(\bm{u}\bm{\cdot} \bm{\nabla})T+\kappa \nabla^{2}T~,
\end{align}
where $\bm{\hat{x}}$ is the east-west direction, $\bm{\hat{y}}$ is the south-north direction, $\bm{\hat{z}}$ is the vertical direction, $\bm{u}$ is the velocity, $\bm{B}$ is the magnetic field, $\bm{\Omega}=\Omega \bm{\hat{e}}_{\Omega}$ is the rotation rate and $\bm{\hat{e}}_{\Omega}=(0,\sin\theta,\cos\theta)$ is the unit vector along $\bm{\Omega}$, $P$ is the pressure, $\rho$ is the mass density of the fluid, $g$ is the gravitational acceleration, $\alpha$ is the coefficient of volume expansion, $\nu$ is the kinematic viscosity, $T$ is the temperature, $\kappa$ is the thermometric conductivity, $\mu$ is the magnetic permeability, and $\eta$ is the magnetic diffusivity. The computational domain is a Cartesian box in the {\it f}-plane and we separate it into two layers, with a convectively stable layer sitting above a convectively unstable layer. In each layer $i$, the horizontal mean temperature gradient $({dT_{0}}/{dz})_{i}$ is assumed be a constant, where $T_{0}$ is the horizontal mean temperature and $i\in \{1,2\}$ is the layer index. We set $({dT_{0}}/{dz})_{1}=-|\beta|$ in the unstable layer and $({dT_{0}}/{dz})_{2}=-\delta|\beta|$ with $\delta\leq 0$ in the stable layer. In such a setting, the parameter $\delta$ measures the relative stability of the upper layer to bottom layer \citep{zhang2002from,cai2020penetrative}. Fig.~\ref{fig:f1} shows the sketch plot of the two-layer structure of our model.

We perform a linear stability analysis of the above equations by decomposing each of the magnetothermal variables into the summation of a horizontal mean and a small perturbation, such as $P=P_{0}+p$, $T=T_{0}+\Theta$, $\bm{B}=\bm{B}_{0}+\bm{b}$. In this paper, we assume that $\bm{B}_{0}$ is a constant vector. As can be shown in the linearized equations, the effects of $x$ and $y$-components of $\bm{B}_{0}$ are similar. For this reason, we further assume that $\bm{B}_{0}$ has a zero component in the $x$-direction. After ignoring the second or higher-order nonlinear terms and choosing the scaling factors
\begin{equation}
    x \to \ell x, t \to \frac{\ell^{2}}{\nu}t, \bm{u} \to \frac{\nu}{\ell}\bm{u}, p \to \frac{\rho \nu^{2}}{\ell^{2}}p, T \to \frac{|\beta|\ell \nu}{\kappa} T, \bm{\bm{b}} \to \frac{\nu}{\eta}|\bm{B}_{0}|\bm{b},
\end{equation}
we can obtain the following linearized equations:
\begin{align}
    &\bm{\nabla} \bm{\cdot} \bm{u}=\bm{\nabla} \bm{\cdot} \bm{b}=0~,\\
    &\partial_{t} \bm{u}= -2E^{-1}\bm{\hat{e}}_{\Omega} \bm{\times} \bm{u}-\bm{\nabla} p +R\Theta \bm{\hat{z}}~\nonumber\\
    &\qquad+\nabla^{2}\bm{u}+E^{-1}\Lambda(\bm{\nabla} \bm{\times} \bm{b}) \bm{\times} \bm{\hat{e}}_{B}~,\label{eq7}\\
    &\partial_{t} \bm{b}=Pm^{-1} \bm{\nabla} \bm{\times} (\bm{u} \bm{\times} \bm{\hat{e}}_{B})+Pm^{-1}\nabla^{2} \bm{b}~,\label{eq8}\\
    &\partial_{t} \Theta=Pr^{-1}\delta_{i}\bm{\hat{z}}\bm{\cdot} \bm{u}+Pr^{-1}\nabla^{2}\Theta~, \label{eq9}
\end{align}
where $\ell$ is the depth of the computational domain, $Pm=\nu / \eta$ is the magnetic Prandtl number, $Pr=\nu / \kappa$ is the Prandtl number, $E=\nu / (\Omega \ell^{2})$ is the Ekman number, $R=(g\alpha |\beta|\ell^{4})/(\nu \kappa)$ is the Rayleigh number, $\Lambda = |\bm{B}_{0}|^{2}/ (\Omega \rho \mu \eta)$ is the Elsasser number, $\bm{\hat{e}}_{B}=(0,\sin\gamma,\cos\gamma)$ is the unit vector along the background magnetic field $\bm{B}_{0}$, $\delta_{i}\in \{1,\delta\}$ is the relative stability parameter in layer $i$. The Elsasser number $\Lambda$ is a dimensionless number that compares the magnetic force to the Coriolis force. It is related to the Chandrasekhar number $Ch$ by $Ch=E^{-1}\Lambda$.

For the convenience of calculation, we decompose the divergence-free fields $\bm{u}$ and $\bm{b}$ into summations of poloidal and toroidal fields by
\begin{align}
    &\bm{u}=\bm{\nabla} \bm{\times} \bm{\nabla} \bm{\times} (\Phi_{u} \bm{\hat{z}})+\bm{\nabla} \bm{\times} (\Psi_{u} \bm{\hat{z}})~,\\
    &\bm{b}=\bm{\nabla} \bm{\times} \bm{\nabla} \bm{\times} (\Phi_{b} \bm{\hat{z}})+\bm{\nabla} \bm{\times} (\Psi_{b} \bm{\hat{z}})~.
\end{align}
After applying $\bm{\hat{z}} \bm{\cdot} \bm{\nabla} \bm{\times}$ and $\bm{\hat{z}}\bm{\cdot} \bm{\nabla} \bm{\times} \bm{\nabla} \bm{\times}$ to the equations (\ref{eq7}) and (\ref{eq8}), we obtain
\begin{align}
    &(D_{z}^{2}+\Delta_{h} -Pm\partial_{t})\Psi_{b}+(\cos\gamma D_{z}+\sin\gamma D_{y}) \Psi_{u}=0~,\\
    &E(D_{z}^{2}+\Delta_{h}-\partial_{t})\Psi_{u}+2(\cos\theta D_{z}+\sin\theta D_{y})\Phi_{u}\nonumber\\
    &\qquad+\Lambda(\cos\gamma D_{z}+\sin\gamma D_{y}) \Psi_{b}=0~,\\
    &(D_{z}^{2}+\Delta_{h}-Pm\partial_{t})\Phi_{b}+(\cos\gamma D_{z}+\sin\gamma D_{y})\Phi_{u}=0~,\\
    &E(D_{z}^{2}+\Delta_{h}-\partial_{t})(D_{z}^{2}+\Delta_{h})\Phi_{u}-2(\cos\theta D_{z}+\sin\theta D_{y})\Psi_{u}~\nonumber\\
    &\qquad-ER\Theta+\Lambda (D_{z}^{2}+\Delta_{h})(\cos\gamma D_{z}+\sin\gamma D_{y}) \Phi_{b}=0~,\\
    &(D_{z}^{2}+\Delta_{h}-Pr\partial_{t})\Theta-\delta_{i}\Delta_{h}\Phi_{u}=0~,
\end{align}
where $\theta$ is the colatitude, $\gamma$ is the inclined angle between $\bm{\Omega}$ and $\bm{B}_{0}$, $D_{z}=\partial / \partial z$, $D_{y}=\partial / \partial y$, and $\Delta_{h}=\partial^2/\partial x^2+\partial^2/\partial y^2$ is the horizontal Laplacian operator. In an $f$-plane, the prognostic variables can be expand as
\begin{align}
    &\{\Phi_{u},\Psi_{u},\Theta,\Phi_{b},\Psi_{b}\}(x,y,z,t)=\nonumber\\
    &\qquad\{\tilde{\Phi}_{u},\tilde{\Psi}_{u},\tilde{\Theta},\tilde{\Phi}_{b},\tilde{\Psi}_{b}\}(z)\exp(ia_{x}x+ia_{y}y-i\sigma t)~,
\end{align}
where $\sigma$ is the time frequency, and $a_{x}$ and $a_{y}$ are the wavenumbers along the $x$ and $y$ directions, respectively. Substituting the expansion series into the linearized equations, we finally obtain the following equations
\begin{align}
    &(D_{z}^{2}-a^{2}+i\sigma Pm)\tilde{\Psi}_{b}+(\cos\gamma D_{z}+i\sin\gamma a_{y})\tilde{\Psi}_{u}=0~,\label{eq18}\\
    &E(D_{z}^{2}-a^{2}+i\sigma)\tilde{\Psi}_{u} + 2(\cos\theta D_{z}+i\sin\theta a_{y})\tilde{\Phi}_{u}\nonumber\\
    &\qquad+\Lambda(\cos\gamma D_{z}+i\sin\gamma a_{y})\tilde{\Psi}_{b}=0~,\label{eq20}\\
    &(D_{z}^{2}-a^{2}+i\sigma Pm)\tilde{\Phi}_{b}+(\cos\gamma D_{z}+i\sin\gamma a_{y})\tilde{\Phi}_{u}=0~,\\
    &E(D_{z}^{2}-a^{2}+i\sigma)(D_{z}^{2}-a^{2})\tilde{\Phi}_{u}-2(\cos\theta D_{z}+i\sin\theta a_{y})\tilde{\Psi}_{u}\nonumber\\
    &\qquad-ER\tilde{\Theta}+\Lambda (D_{z}^{2}-a^{2})(\cos\gamma D_{z}+i\sin\gamma a_{y})\tilde{\Phi}_{b}=0~,\\
    &(D_{z}^{2}-a^{2}+i\sigma Pr)\tilde{\Theta}+\delta_{i}a^{2}\tilde{\Phi}_{u}=0~,\label{eq22}
\end{align}
where $a^{2}=a_{x}^{2}+a_{y}^{2}$.

To solve the above equations, twelve boundary conditions are required for the prognostic variables. We set impenetrable, stress-free, and isothermal at both the top and bottom boundaries. It requires
\begin{align}
    &\tilde{\Phi}_{u}(0)=D_{z}^{2}\tilde{\Phi}_{u}(0)=D_{z}\tilde{\Psi}_{u}(0)=\tilde{\Theta}(0)=0~,\\
    &\tilde{\Phi}_{u}(1)=D_{z}^{2}\tilde{\Phi}_{u}(1)=D_{z}\tilde{\Psi}_{u}(1)=\tilde{\Theta}(1)=0~.
\end{align}
For the magnetic field, we assume electrically non-conducting boundary conditions \citep{Cbook,jones2000Convection} at both the top and bottom boundaries with
\begin{align}
    & \tilde{\Psi}_{b}(0)=\tilde{\Psi}_{b}(1)=0~,\\
    & D_{z}\tilde{\Phi}_{b}(0)=a\tilde{\Phi}_{b}(0)~,\\
    & D_{z}\tilde{\Phi}_{b}(1)=-a\tilde{\Phi}_{b}(1)~.
\end{align}
Given these boundary conditions and the parameters $\gamma$, $\theta$, $Pm$, $Pr$, $E$, $\lambda$, $\delta$ and $R$, Eqs.~(\ref{eq18}-\ref{eq22}) forms an eigenvalue problem for $\sigma$. In our computation, we fix $\gamma$, $\theta$, $Pm$, $Pr$, $E$, $\Lambda$, $\delta$ and vary $R$, therefore $\sigma=\sigma(R|\gamma,\theta, Pm, Pr, E, \Lambda, \delta)$. Let $\sigma=\sigma_{r}+i\sigma_{i}$ be a complex number, where $\sigma_{r}$ and $\sigma_{i}$ are the real and imaginary parts, respectively. The flow is stable when $\sigma_{i}<0$ and unstable when $\sigma_{i}>0$, thus the transition state for the onset of convection occurs at $\sigma_{i}=0$. We discretize Eqs.(\ref{eq18}-\ref{eq22}) by a staggered grid finite difference scheme, and numerically solve the eigenvalue problems of $\sigma$. By gradually increasing $R$, we numerically search for the critical value of $R$ at the transition state for the onset of convection ($\sigma_{i}=0$). If $\sigma_{r}=0$ at the transition state, then the onset of convection is stationary; otherwise it is oscillatory.

The convective flux $F$ has been widely used as a proxy of measuring overshooting or penetrative distances in stars \citep{zahn1991convective,browning2004simulations,kapyla2017extended}. In an incompressible flow, the horizontally averaged value of the convective flux can be computed as
\begin{align}
\langle F \rangle = \langle \Re(w)\Re(\Theta) \rangle=\frac{1}{2}a^2 \Re(\tilde{\Phi}_{u}\tilde{\Theta}^{*})~,\label{eq28}
\end{align}
where the operator $\Re$ represents the real part of a complex number, the symbol star denotes the complex conjugate, and $w$ is the vertical component of $\bm{u}$. With the value of $\langle F \rangle$, the penetrative distance can be defined at the location where the convective flux has attained a certain fraction (we choose 5\% in our calculation) of its most negative value in the convectively stable layer \citep{browning2004simulations}. An example on calibrating penetrative distance is given at Appendix~\ref{appendixA}.

\section{The asymptotic analysis}
We focus on fast-rotating stars or planets, where flow patterns are most likely aligned as Taylor-Proudman columns. Under this assumption, we can set the wavenumber ratio as $a_{y}/a=\cos\theta/\sqrt{1+\cos^2\theta}$. The Taylor-Proudman constraint can be understood by taking a curl of Eq.~(\ref{eq7}), which yields
\begin{align}
    &2^{-1}E\bm{\nabla} \bm{\times}(\partial_{t} \bm{u}-R\Theta \bm{\hat{z}}-\nabla^{2}\bm{u})= (\bm{\hat{e}}_{\Omega} \bm{\cdot} \bm{\nabla})\bm{u} ~\nonumber\\
    &\qquad+2^{-1}\Lambda(\bm{\hat{e}}_{B} \bm{\cdot} \bm{\nabla})(\bm{\nabla} \bm{\times} \bm{b})~.
\end{align}
If both $E$ and $\Lambda$ are small (with rapid rotation and weak background magnetic field), then to leading order the above equation can be reduced into $(\bm{\hat{e}}_{\Omega} \bm{\cdot} \bm{\nabla})\bm{u}\approx 0$. Therefore the velocity remains almost a constant along the rotational axis. However, even with a small $E$ (rapid rotation), the Taylor-Proundman constraint can be broken if $\Lambda$ is large (strong background magnetic field). In such a case, the leading order will be $(\bm{\hat{e}}_{B} \bm{\cdot} \bm{\nabla})(\bm{\nabla} \bm{\times} \bm{b})\approx 0$, which means that the current density remains almost a constant along the direction of $\bm{\hat{e}}_{B}$. When $Pm$ is small, it can be found from Eq.(\ref{eq8}) that $\bm{\nabla} \bm{\times} \bm{b}$ has the same order as $\bm{u}$. Thus we expect that the Taylor-Proudman constraint is broken when $\Lambda \sim 1$. If $\bm{B_{0}}$ has the same direction as $\bm{\Omega}$, then the columnar structure is still expected to be hold even when $\Lambda>1$.

In this paper, we mainly focus on the penetrative convection in planetary-like conditions with $E\ll 1$ (rapid rotation), under which the penetrative distance can be qualitatively estimated. We follow the idea proposed by \citet{takehiro2001penetration} and \citet{takehiro2015penetration} to analytically estimate the distance of penetration. If $\Lambda \ll 1$, then the magnetic effect is negligible and the problem degenerates into penetrative convection in rapidly rotation. Ignoring the diffusive terms, we can obtain the wave equation in each layer (\cite{gerkema2005near,cai2021Inertial})
\begin{equation}
\nabla^2 w_{tt}+4E^{-2}(\bm{\hat{e}}_{\Omega}\bm{\cdot}\bm{\nabla})^2 w-RPr^{-1}\delta_{i}\nabla_{H}^2 w=0~, \label{eq30}
\end{equation}
from which the vertical wavenumber $k$ satisfies
\begin{equation}
(a^2+k^2)\sigma^2-4E^{-2}(\sin\theta a_{y}+\cos\theta k)^2+RPr^{-1}\delta_{i}a^2=0~. \label{eq31}
\end{equation}
The solution is wave-like when $k^2>0$, while it is evanescent when $k^2<0$. It can be shown that the existence of wave-like solution requires \citep{cai2021Inertial}
\begin{equation}
\sigma^4-(f^2+f'^2\sin^2\alpha+N_{i}^2)\sigma^2+N_{i}^2 f^2<0~, \label{eq32}
\end{equation}
where $\sin\alpha=a_{y}/a$, $f=2E^{-1}\cos\theta$ is the traditional Coriolis parameter, $f'=2E^{-1}\sin\theta$ is the non-traditional Coriolis parameter, and $N_{i}^2=-RPr^{-1}\delta_{i}$ is the square of Brunt-V\"as\"al\"a frequency. Obviously, $N_{i}^2$ is negative in the unstable layer, but positive in the stable layer. Therefore, it is possible for inertial wave to propagate in the unstable layer, and gravito-inertial wave to propagate in the stable layer. At the low $Pr$ flow, the onset of convection usually occurs firstly as oscillatory convection ($\sigma^2 > 0$). Let us at first consider a simple case in a non-tilted $f$-plane at $\theta=0$. In such a case, it requires $\sigma^2 \in (0,f^2)$ for the inertial wave to propagate in the unstable layer, and $\sigma^2 \in (\min(N_{2}^2,f^2),\max(N_{2}^2,f^2))$ for the gravito-inertial wave to propagate in the stable layer. Here we note that wave solutions can exist in both the unstable and stable layers when $N_{2}^2 < f^2$ (stable layer is weakly stratified). Otherwise, when $N_{2}^2>f^2$, wave is evanescent in the stable layer. In the tilted $f$-plane, the situation is similar. In the following, we examine these two possibilities in the context of weakly and strongly stratified layers at the top.

For the weakly stratified case, \citet{cai2021Inertial} have shown that in a tilted $f$-plane, waves can be efficiently transmitted from the unstable layer to stable layer when $|N_{2}^2-N_{1}^2|\ll 4E^{-2}$ or $\sigma^2 \rightarrow f^2$. If we define $Ro=(RPr^{-1})^{1/2}E$ as the Rossby number, then the first condition $|N_{2}^2-N_{1}^2|\ll 4E^{-2}$ is equivalent to $Ro \ll 2(1+\delta)^{-1/2}$. For the rapidly rotating convection at small $Pr$, the asymptotic solution shows the critical value of Rayleigh number $R\sim O(Pr^{4/3}E^{-4/3})$ \citep{Cbook}. It gives $Ro \sim O(Pr^{1/6}E^{1/3})$, so that the first condition can be satisfied when $Pr$ and $E$ are small. In other words, the penetration can be efficient at small $Pr$ and $E$. The second condition $\sigma^2 \rightarrow f^2$ is satisfied when the waves propagate at the critical latitudes $\cos\theta =\pm \sigma E/2$. From the asymptotic solution $\sigma \sim O(Pr^{-1/3}E^{-2/3})$ \citep{Cbook}, we can deduce $\sigma E \sim Pr^{-1/3}E^{1/3}$, which approaches to zero when $E\rightarrow 0$. Apparently, the penetration can be efficient at low latitudes when $E$ is small. The efficient penetrative convection at small $Pr$, small $E$, and small $|\delta|$ have been verified in \citet{zhang2000teleconvection} for spherical geometry, and in \citet{cai2020penetrative} for tilted $f$-plane at low latitudes. For efficient wave transmission, the wavenumbers, as calculated by Eq.~\ref{eq31}, exhibit near equivalence in both unstable and stable layers, thereby leading to resonance. As a result, the penetration is also efficient. One special case of penetrative convection is the appearance of teleconvection, which usually occurs at small $Pr$ and small $|\delta|$ \citep{zhang2002from}.
From Eq.(\ref{eq9}), it is obvious that the relation $\Theta \sim O(|\delta_{i}| w)$ holds in the limit $Pr\rightarrow 0$. When $\delta_{1}=1$ and $\delta_{2}\rightarrow 0$, to balance the wave fluxes in both sides, the magnitude of $w$ in the stable layer must be much greater than that in the unstable layer. As a result, it drives more vigorous fluid motions in the stable than unstable layers. This explains the phenomena teleconvection as observed in spherical geometry \citep{zhang2000teleconvection,zhang2002from} and Cartesian geometry \citep{cai2020penetrative}. Despite the vigorous motion observed in the stable layer, it's worth noting that the convective flux may remain small due to the slight temperature perturbation.

For strongly stratified case ($|\delta|\gg 1$), $\sigma^2$ is small compared to $f^2$ and $N_{2}^2$.
Ignoring the first term in Eq.(\ref{eq31}) leads to
\begin{equation}
k \approx (2\cos\theta)^{-1}  Ro \delta_{i}^{1/2} a-\sin\theta (1+\cos^2 \theta)^{-1/2}a~,
\end{equation}
for $\theta \neq \pi/2$.
As $\delta_{2}$ is negative, the wave is evanescent in the stable layer. The imaginary part of $k$ gives the decaying rate of the evanescent wave as $\Im(k)=(2\cos\theta)^{-1} a Ro |\delta|^{1/2}$. Then the e-folding penetrative distance can be estimated as $\Delta = 2\cos\theta (a Ro  |\delta|^{1/2} )^{-1} $. For onset of rapidly rotating convection at $\theta=0$, the analysis shows that $R \sim O(E^{-4/3})$ and $a \sim O(E^{-1/3})$ when $E\rightarrow 0$ \citep{Cbook}, which yields $a Ro \sim O(1)$. In the tilted $f$-plane, appropriate scalings will be $R \sim O( (2\cos\theta)^{4/3} E^{-4/3})$ and $a \sim O((2\cos\theta)^{1/3}E^{-1/3})$, which gives $a Ro \sim O(2\cos\theta)$. Therefore the penetrative distance has a scaling $\Delta \sim O(|\delta|^{-1/2})$. We note that this relation does not depend on $E$ when $E \rightarrow 0$. For the onset of convection, the penetrative distance is largely dependent on the degree of stratification in the upper layer. It is anticipated that a highly stratified stable layer tends to inhibit the penetration.

If $\Lambda \gg 1$, then the magnetic effect must be included. Again, by ignoring the diffusive terms, we can obtain the wave equations for magneto-convection as
\begin{align}
&[\partial_{tt}-ChPm^{-1}(\bm{\hat{e}}_{B}\cdot \bm{\nabla})^2]^2\nabla^2 w+4E^{-2}(\bm{\hat{e}}_{\Omega}\bm{\cdot}\bm{\nabla})^2 \partial_{tt}w \nonumber\\
&-[\partial_{tt}-ChPm^{-1}(\bm{\hat{e}}_{B}\cdot \bm{\nabla})^2]RPr^{-1}\delta_{i} \nabla_{h}^2 w=0~.\label{eq34}
\end{align}
When $\Lambda \gg 1$, the Coriolis term becomes negligible. If a wave solution is present in both the unstable and stable layers, it necessitates a balance between the first and third terms on the left hand side. This yields
\begin{align}
&[\partial_{tt}-ChPm^{-1}(\bm{\hat{e}}_{B}\cdot \bm{\nabla})^2][\partial_{tt}\nabla^2w-ChPm^{-1}(\bm{\hat{e}}_{B}\cdot \bm{\nabla})^2\nabla^2 w \nonumber\\
&-RPr^{-1}\delta_{i} \nabla_{h}^2 w]=0~.\label{eq35}
\end{align}
The first factor in the above equation gives the solution for pure Alfv\'en wave, which does not depend on $\delta_{i}$.
Our analysis will be confined to wave solutions that are derived from the second factor, which encompasses internal gravity waves (IGWs) and slow magnetic wave (SMWs). If the operator $\nabla^2$ is replaced with $-(a^2+k^2)$, the wave equation for the second factor transforms into
\begin{equation}
\nabla^2w_{tt}+ChPm^{-1}(k^2+a^2)(\bm{\hat{e}}_{B}\cdot \bm{\nabla})^2 w -RPr^{-1}\delta_{i} \nabla_{h}^2 w=0~.\label{eq36}
\end{equation}
which bears a resemblance to Eq.(\ref{eq30}). Now we examine the scenario where wave solutions are present in both the unstable and stable layers. Wave transmission can be efficient when the waves in both layers have similar vertical wavenumbers (resonance). Analogous to Eq.(\ref{eq30}), we find that resonance is likely when the value of $RCh^{-1}Rb(1+\delta)$ is small (or $RCh^{-1}Rb\ll (1+\delta)^{-1}$), where $Rb=PmPr^{-1}$ is the Roberts number. When $\gamma=0$ and $Ch \rightarrow \infty$, asymptotic analysis shows $R\sim O(Rb^{-2}Ch)$ for the onset of oscillatory convection and $R\sim O(Ch)$ for onset of stationary convection \citep{Cbook}. As a result, $RCh^{-1}Rb$ has an order of $O(Rb^{-1})$ for oscillatory convection and $O(1)$ for stationary convection. Therefore, efficient penetration is more probable when $\max(Rb^{-1},1) \ll (1+\delta)^{-1}$. This implies that the relative stratification $\delta$ needs to be weak for efficient penetration.

Now we turn to the case when wave is evanescent in the stable layer. Given our interest in slow waves, we can disregard the time derivative term in the second factor. This leads to the following
\begin{align}
(a_{y}\sin\gamma+k\cos\gamma )^2 (k^2+a^2) \approx Ch^{-1}Rb R\delta_{i} a^2 ~.\label{eq37}
\end{align}
Note that it is also valid for stationary convection.
As short waves cannot penetrate too far, we can ignore higher order terms of $k$ in the above equation, which yields
\begin{align}
&[\cos^2\gamma+(a_{y}/a)^2\sin^2\gamma]k^2+ (a_{y} \sin2\gamma) k \nonumber\\
&+a_{y}^2\sin^2 \gamma -Ch^{-1} Rb R\delta_{i}\approx 0~.
\end{align}
Therefore, the imaginary part of $k$ can be approximated as
\begin{align}
\Im(k)\approx \left[\frac{Ch^{-1}Rb R|\delta|}{\cos^2\gamma+(a_{y}/a)^2\sin^2\gamma}\right]^{1/2}~.
\end{align}
In a tilted $f$-plane, a suitable scaling that accounts for the effect of inclination can be represented as $R\sim O(Rb^{-2}Ch\cos^2\gamma)$ for oscillatory convection, and $R\sim O(Rb^{-1}Ch\cos^2\gamma)$ for stationary convection. It is not easy to distinguish oscillatory convection from stationary convection in the analysis. Here we just ignore the contribution from $Rb$, which gives
\begin{align}
\Im(k)\sim O\left[\frac{|\delta|\cos^2 \gamma}{\cos^2\gamma+(a_{y}/a)^2\sin^2\gamma}\right]^{1/2}~. \label{eq39}
\end{align}
At $\gamma=\theta$ or $\gamma=0$, we obtain $\Im(k)\sim O(|\delta|^{1/2})$, which results in a penetrative distance of $\Delta \sim O(|\delta|^{-1/2})$.  In both magnetism-dominated and rotation-dominated flow, the scaling of penetrative distance on the intensity of stratification is the same.

\section{The numerical result}
We numerically investigate the effect of magnetic field on penetrative distances in rotating convection. As we have several parameters, it is too expensive to explore the full parameter space. To reduce the complexity of the problem, we set $Pr=0.1$ and $Pm=0.1$, which are typical values in stars and planets \citep{schubert2011planetary}. We investigate the problem of penetrative magneto-convection problem by varying the values of $E$ from $10^{-7}$ to $10^{-3}$, $\theta$ from $0$ to $\pi/2$, and $\delta$ from $-15$ to $-0.1$.  We examine the effects of weak, moderate, and strong imposed magnetic fields by setting $Ch$ to $1$, $10^3$, and $10^6$, respectively. The background magnetic field $\bm{B}_{0}$ can be imposed at different angle. We mainly consider two situations: one is that $\bm{B}_{0}$ is parallel to $\bm{\Omega}$ ($\gamma=\theta$), and the other is that $\bm{B}_{0}$ is parallel to $\bm{\hat{z}}$ ($\gamma=0$).

\subsection{Weakly stratified stable layer}
\begin{figure}
    \centering
    \subfigure[]{
    \includegraphics[width=0.45\columnwidth]{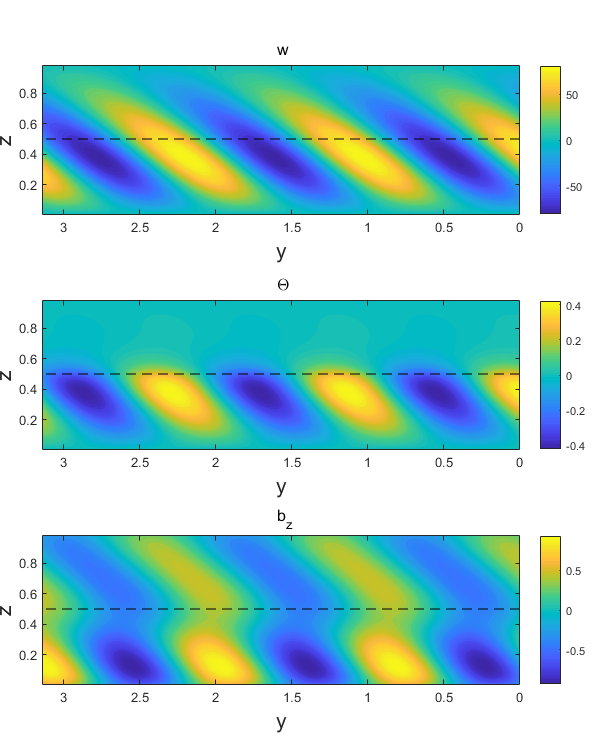}}
    \subfigure[]{
    \includegraphics[width=0.45\columnwidth]{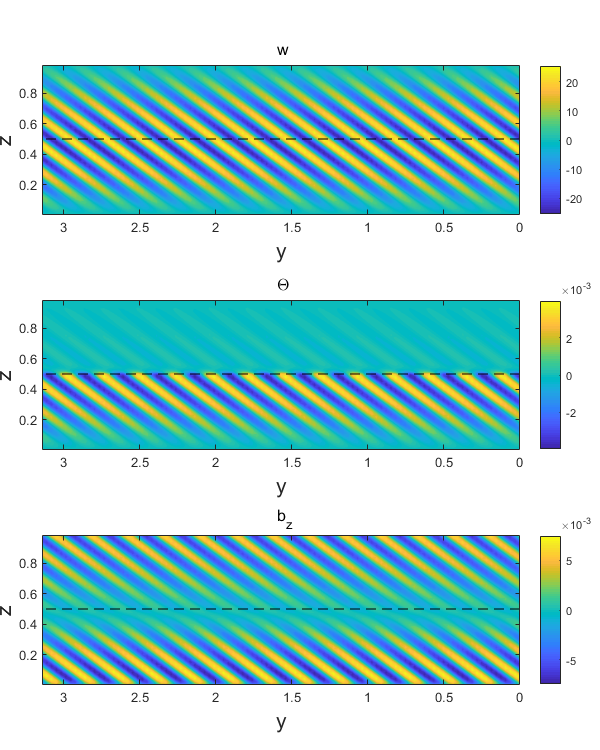}}
    \subfigure[]{
    \includegraphics[width=0.45\columnwidth]{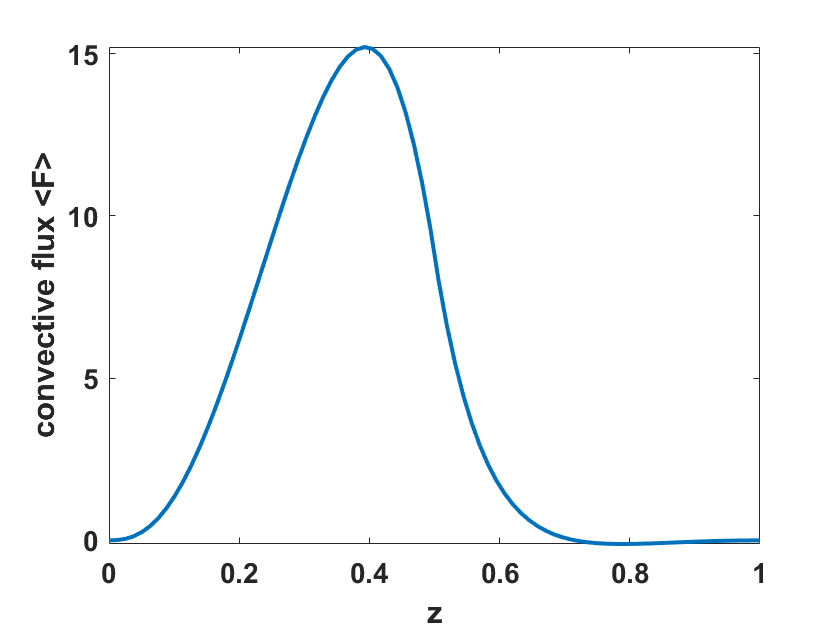}}
    \subfigure[]{
    \includegraphics[width=0.45\columnwidth]{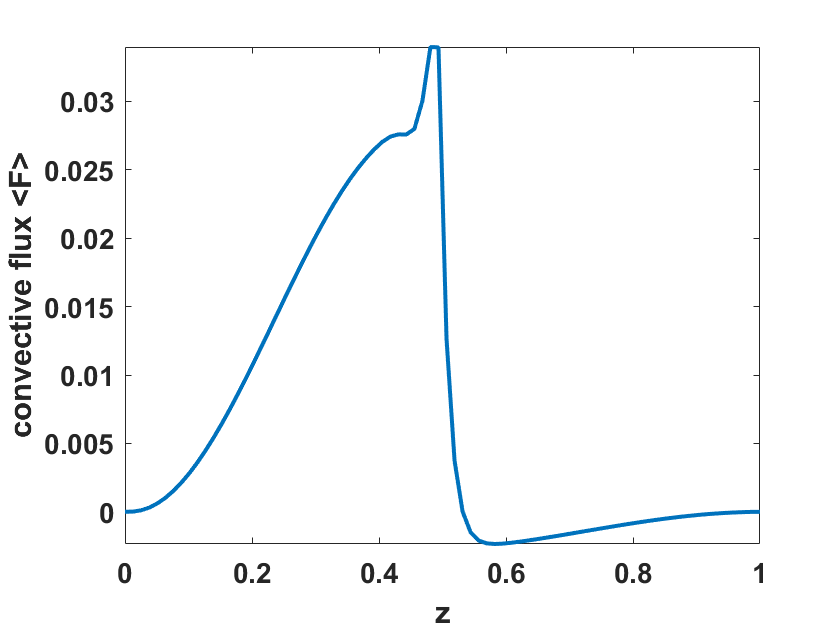}}
    \caption{The contour plots for the vertical velocity $w$, the temperature perturbation $\Theta$, and the vertical component of magnetic perturbation $b_{z}$, at $\delta=-0.1$, $\theta=0.3\pi$, $Pr=Pm=0.1$. The left and right columns use different $E$ and $Ch$ with (a)$E=10^{-3}$, $Ch=10^{6}$ and (b)$E=10^{-6}$, $Ch=1$, respectively. (c) and (d)The corresponding convective flux $\langle F\rangle$ as a function of $z$.}
    \label{fig:f2}
\end{figure}

For a weakly stratified stable layer, our computation indicates that the penetration is typically highly efficient. Therefore, quantifying the penetrative distance becomes irrelevant, as the flow can permeate the entirety of the stable layer. Here we use two cases to illustrate that the effective penetration for a weakly stratified stable layer. Both cases choose $Pr=Pm=0.1$, $\delta=0.1$, and $\gamma=\theta=0.3\pi$. The first case is characterized by a strong background magnetic field and a slow rotation rate, with $Ch=10^6$ and $E=10^{-3}$ (magnetism-dominated flow). In contrast, the second case features a weak background magnetic field and a fast ration rate, with $Ch=1$ and $E=10^{-6}$ (rotation-dominated flow).
Fig.~\ref{fig:f2} displays the contour plots of $w$, $\Theta$, and $b_{z}$ for these two cases. In each case, $w$ aligns with the direction of rotation and extends to the top of the box. Conversely, the temperature perturbation is primarily confined to the unstable layer. As previously stated, $\Theta \sim O(|\delta_{i}| w)$ when $Pr$ is small. The small $\delta$ results in a weak temperature perturbation in the stable layer. In both cases, $b_{z}$ also reaches the top of the box. However, their pattern exhibit significant differences. In the first case, $b_{z}$ displays a phase shift from the unstable layer to the stable layer. In contrast, in the second case, $b_{z}$ maintains nearly the same phase from the unstable to stable layers. From Eq.~(\ref{eq20}), it is evident that $b_{z}\propto (\cos\gamma D_{z}+\sin\gamma D_{y})w$. The phase shift of $b_{z}$ is contingent on which term on the right-hand side is dominant. If $\cos\gamma D_{z}$ is the dominant term, a phase shift is anticipated. Conversely, if $\sin\gamma D_{y}$ dominates, $b_{z}$ is expected to maintain the same phase. Clearly, the latter necessitates a strong nontraditional effect.

Now we examine the convective flux for these two cases. In the case of penetrative convection, the convective flux $\langle F\rangle$ typically changes sign at the interface between the unstable and stable layers, given that $\Theta$ is anticipated to switch sign at this interface. However, the situation may differ if wave solutions can persist in both the unstable and stable layers. As demonstrated in \citet{cai2021Inertial}, under certain conditions, the convective flux can maintain near-constant values across the interface. Figs.~\ref{fig:f2}(c) and (d) show the convective fluxes for the two cases, respectively. In Fig.~\ref{fig:f2}(c) for the magnetism-dominated case, we note that the convective flux remains positive over a considerable distance within the stable layer, extending away from the interface. In this case, the numerical result gives $\sigma=142.3$ and $a=10.7$ for the onset of convection. If we ignore the rotational effect and solve the Eq.~\ref{eq36}, we discover that Alfv\'en wave can propagate in the unstable layer, while IGW and SMW can survive in the stable layer. Consequently, it's feasible for waves to sustain in both unstable and stable layers, resulting in efficient penetration. In Fig.~\ref{fig:f2}(d), which represents the rotation-dominated case, the convective flux exhibits a rapid decline across the interface transitioning from the unstable to stable layers. Even though the value becomes negative, its magnitude experiences a gradual decay moving away from the interface within the stable layer.  In this instance, the result gives $\sigma=-5468.8$ and $a=52.1$ at the onset of convection. Once again, we observe that the inertial wave can persist in the unstable layer, and the gravito-inertial wave can survive in the stable layer.  Consequently, the penetration can be efficient since wave solutions are allowed in both the unstable and stable layers.

\subsection{Strongly stratified stable layer}
\subsubsection{$\mathbf{B}_{0}$ is parallel to $\mathbf{\Omega}$}
\begin{figure}
    \centering
    \subfigure[]{
    \includegraphics[width=0.45\columnwidth]{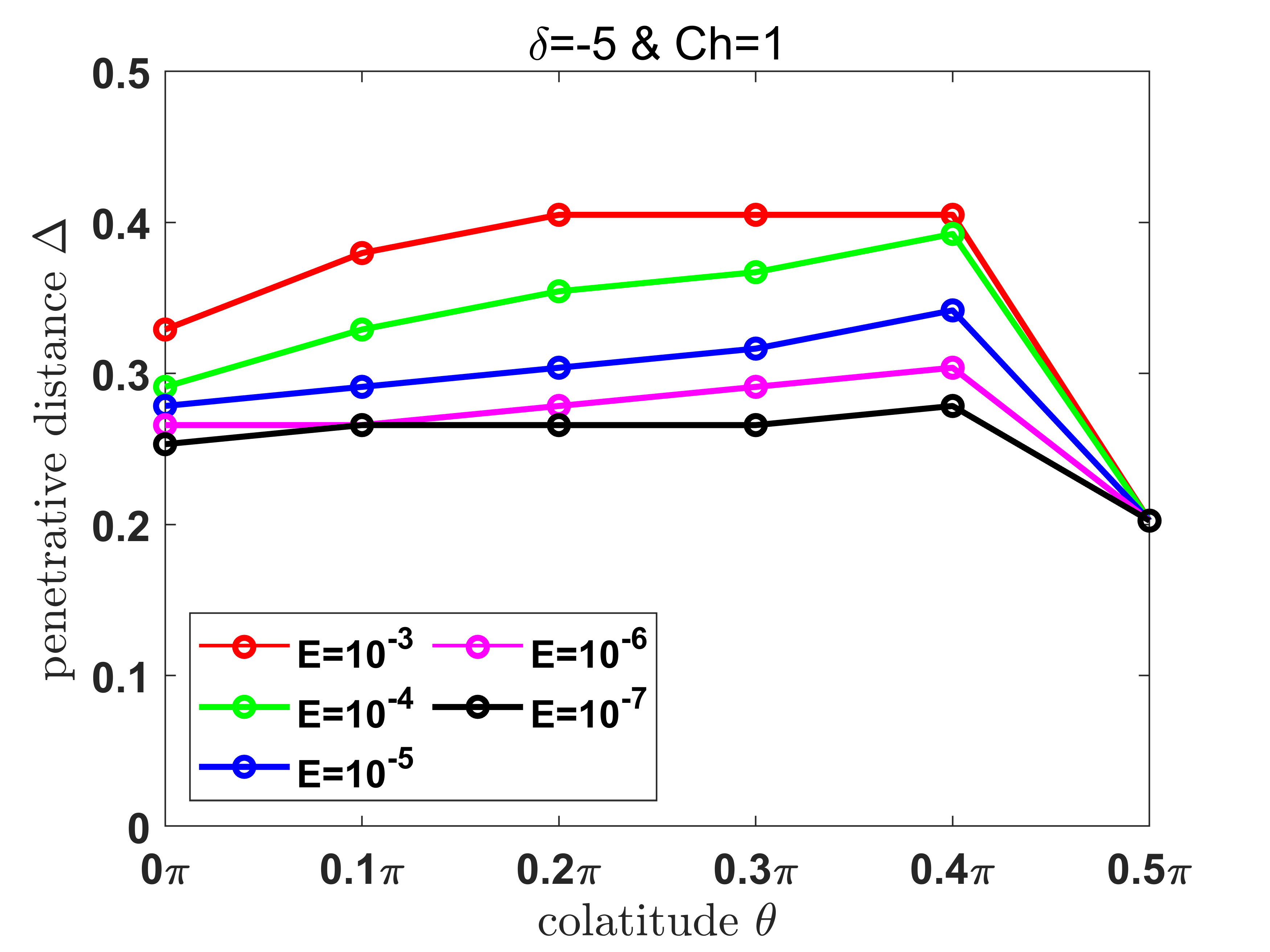}}
    \subfigure[]{
    \includegraphics[width=0.45\columnwidth]{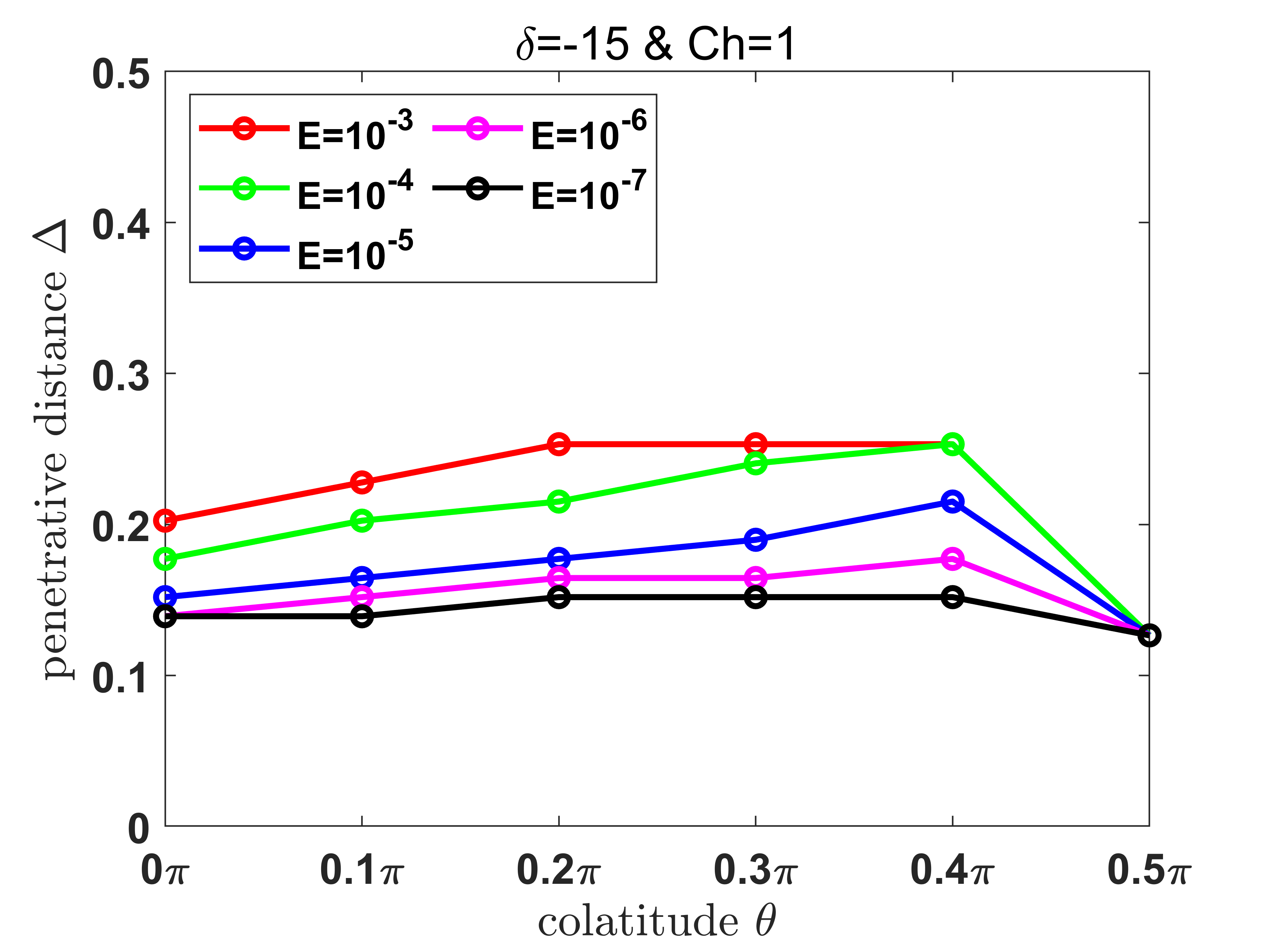}}
    \subfigure[]{
    \includegraphics[width=0.45\columnwidth]{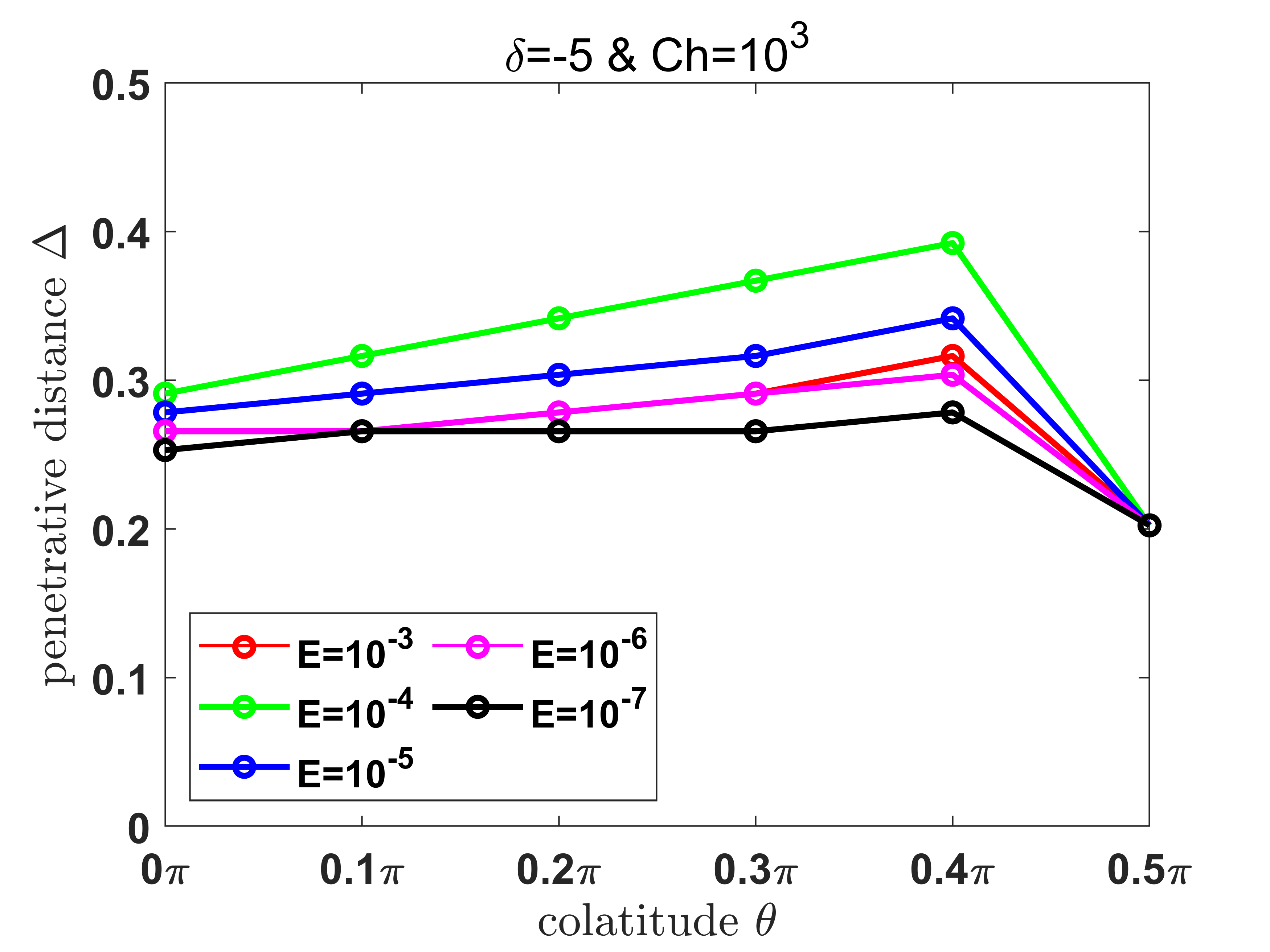}}
    \subfigure[]{
    \includegraphics[width=0.45\columnwidth]{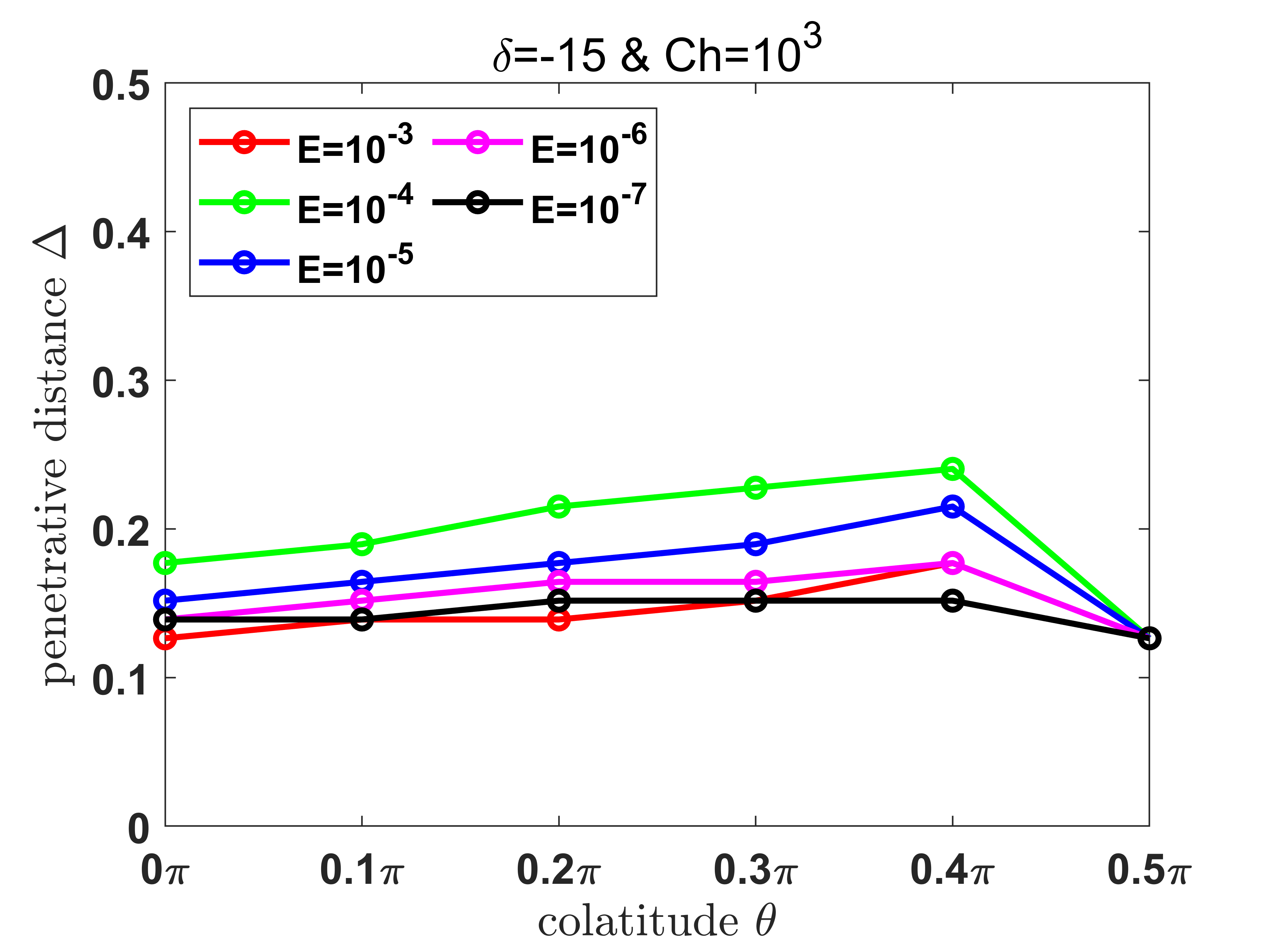}}
    \subfigure[]{
    \includegraphics[width=0.45\columnwidth]{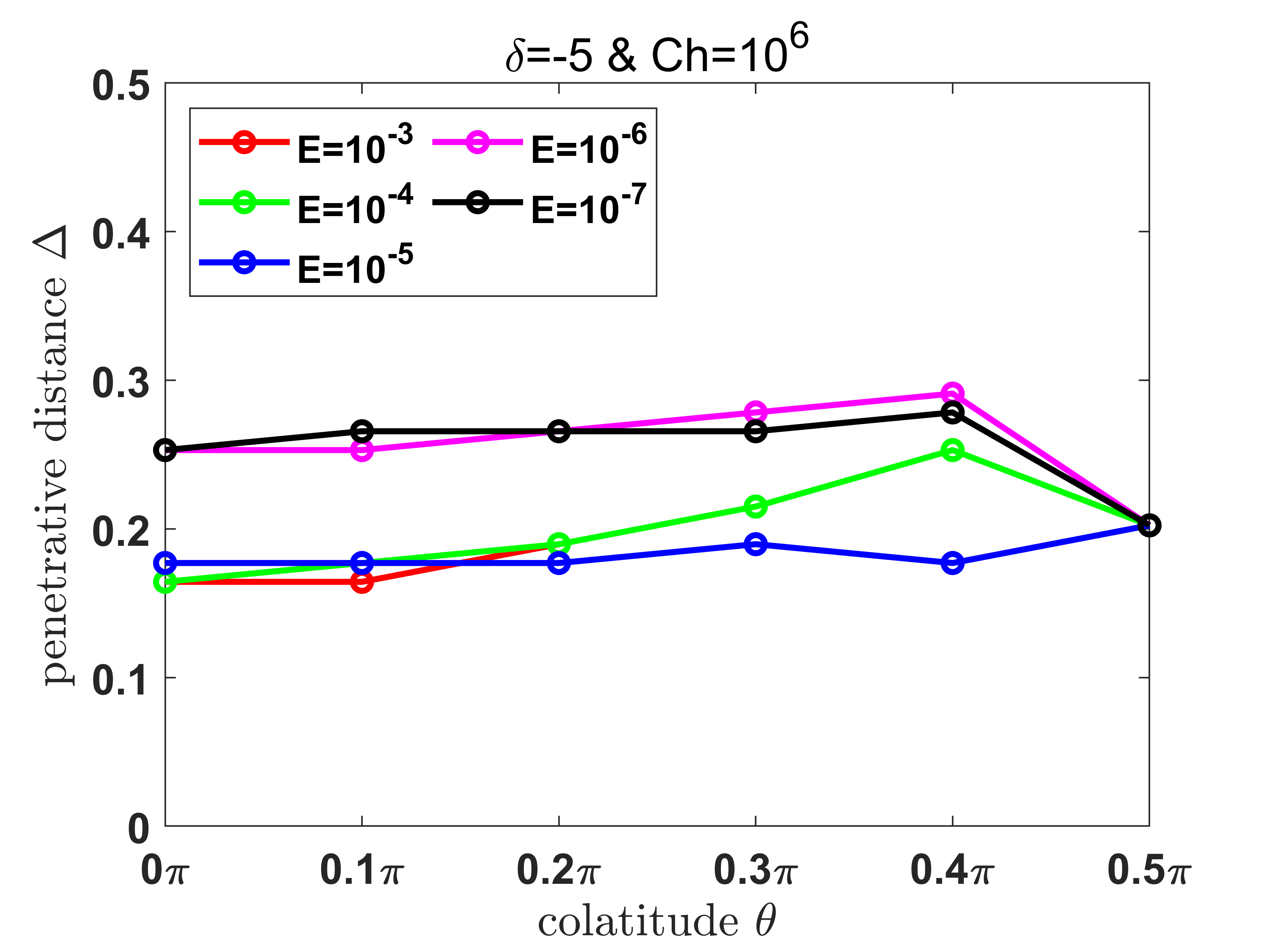}}
    \subfigure[]{
    \includegraphics[width=0.45\columnwidth]{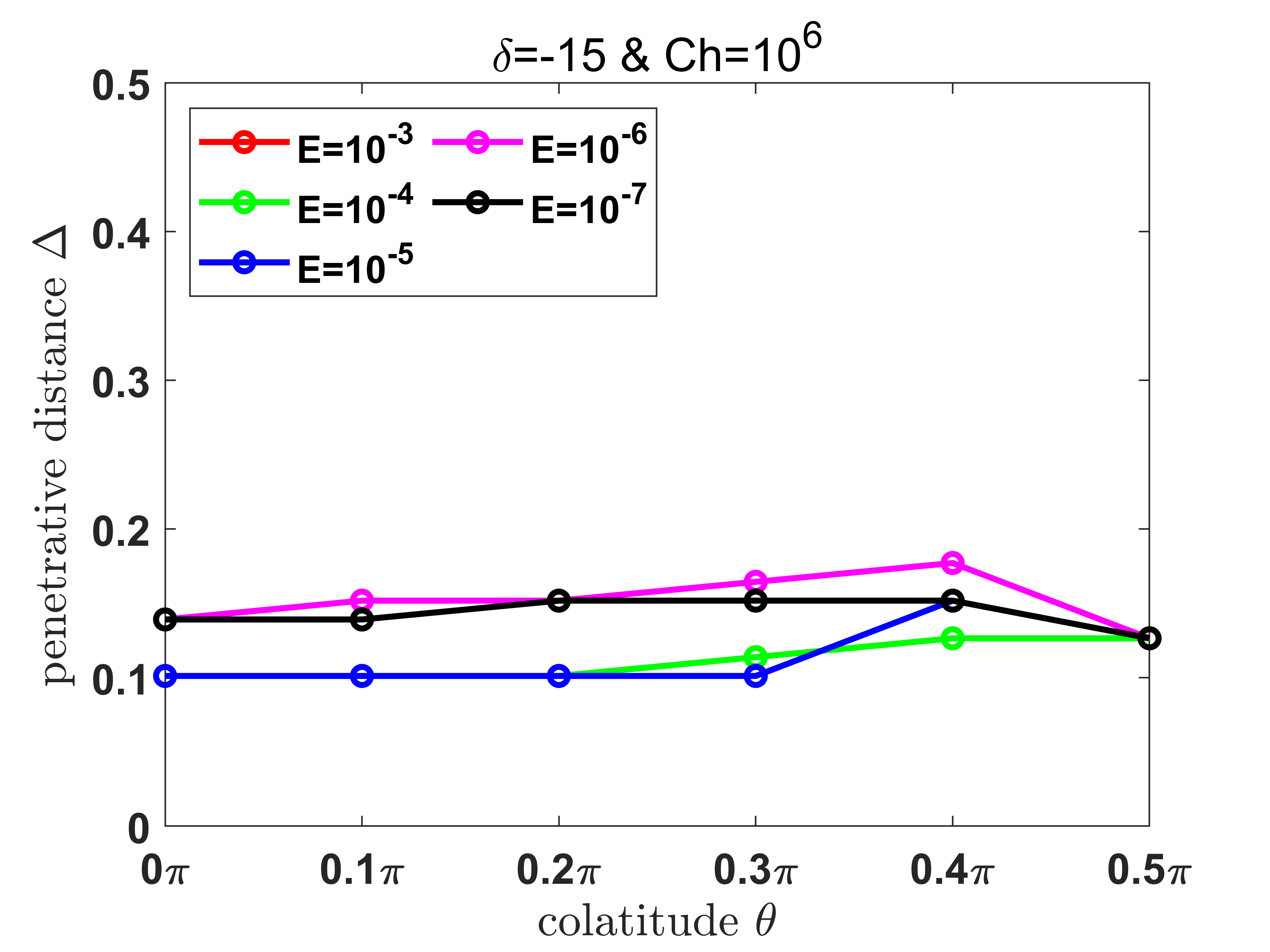}}
    \caption{Penetrative distances for cases $\delta=-5$ or $\delta=-15$ when $\mathbf{B}_{0}$ is parallel to $\mathbf{\Omega}$. Panels (a)-(b), (c)-(d), (e)-(f) show the results of $Ch=1$, $Ch=10^{3}$ and $Ch=10^{6}$, respectively.}
    \label{fig:f3}
\end{figure}

When the vector $\mathbf{B}_{0}$ is parallel to $\mathbf{\Omega}$, it follows that $\gamma=\theta$. Fig.~\ref{fig:f3} illustrates the dependence of the penetrative distance $\Delta$ on the parameters $E$, $Ch$, and $\delta$. The penetrative distance is quantified using the convective flux $\langle F \rangle$ as a proxy. It is evident that stratification significantly influences the penetrative distance. The penetrative distances for $\delta=-15$ are considerably shorter than those for $\delta=-5$. This is expected as stronger stratification tends to suppress convection, resulting in a reduced penetrative distance. The numerical results reveal that the penetrative distances for $\delta=-5$ are approximately $1.7$ times greater than those for $\delta=-15$ when $E=10^{-7}$. This finding aligns well with the asymptotic relation $\Delta \sim O(\delta^{-1/2})$ derived from our prior analysis for $E\rightarrow 0$. When $E>10^{-7}$, the penetrative distance exhibits a slight dependence on latitude. The prevailing pattern indicates a reduction in the depth of penetration from low latitudes to high latitudes. In this discussion of the trend, the results at the equator are omitted, as the onset of convection at the equator is always stationary. The stratification's impact on the depth of penetration is more pronounced at higher latitudes compared to lower ones, indicating the significant role of the non-traditional Coriolis force in penetrative magneto-convection.

Apart from stratification, both rotation and magnetic field have prominent effects on the penetrative distance. The rotational effect on penetration varies with the background magnetic strength. For a weak magnetic field, we see that $\Delta$ almost decreases with decreasing $E$ monotonically at all latitudes (see the first row of Fig.~\ref{fig:f3}). It means that rotation basically has a negative effect on penetration. On the other hand, for a strong magnetic field (see the third row of Fig.~\ref{fig:f3}), we observe that the penetrative distance first decreases with decreasing $E$ in the range of $(10^{-5},10^{-3})$. However, when $E$ further decreases to $(10^{-7},10^{-6})$, the penetrative distance increases abruptly and reaches an equilibrium. The deviation in these two groups is probably due to the change of the relative strength of magnetic to rotational effects. When $Ch=10^6$, the Elsasser number is $\Lambda> 1$ for $E\in [10^{-5},10^{-3}]$, which indicates a stronger magnetic effect than rotational effect. As the rotational effect is less important, the penetrative distance is mainly affected by the magnetic effect. In Figs.~\ref{fig:f3} (e) and (f), the curves of $E=10^{-3}$ overlaps those of $E=10^{-4}$. It indicates that the rotational effect is almost negligible if $\Lambda \gg 1$. Also, the penetrative distance shows an increasing trend from high to low latitudes. In the contrast, for $E\in [10^{-7},10^{-6}]$, the Elsasser number is $\Lambda\leq 1$. In such cases, the penetrative distance is mainly determined by the dominating rotational effect. It seems that the penetrative distance dominated by rotation will differ from that dominated by magnetism by a factor of $\sqrt{2}$ at $\theta=0$. The difference has been clearly shown in Figs.~\ref{fig:f3}(e) and (f) for the two groups  $E\in [10^{-7},10^{-6}]$ and $E\in [10^{-5},10^{-3}]$.

\begin{figure}
    \centering
    \subfigure[]{
    \includegraphics[width=0.45\columnwidth]{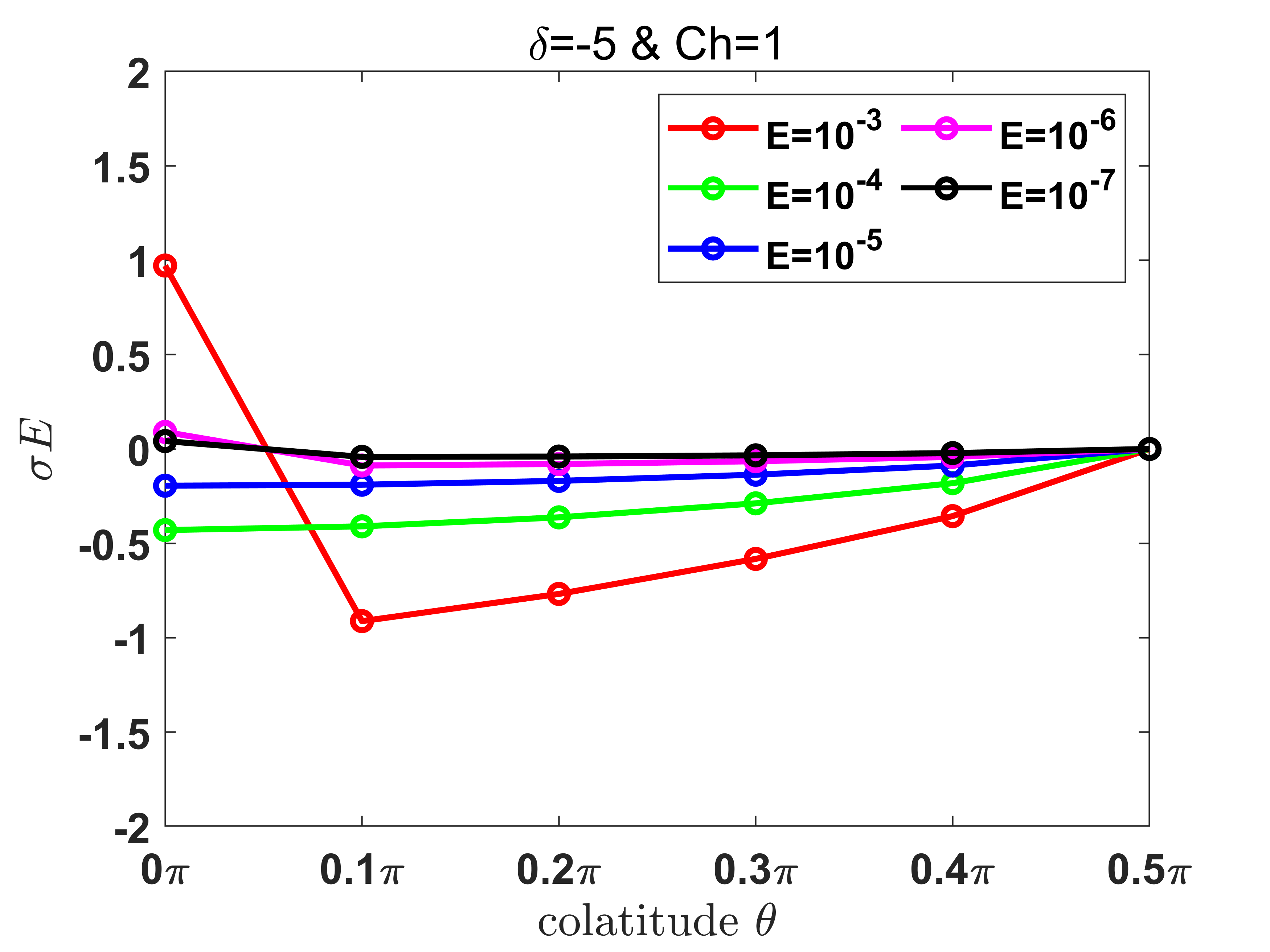}}
    \subfigure[]{
    \includegraphics[width=0.45\columnwidth]{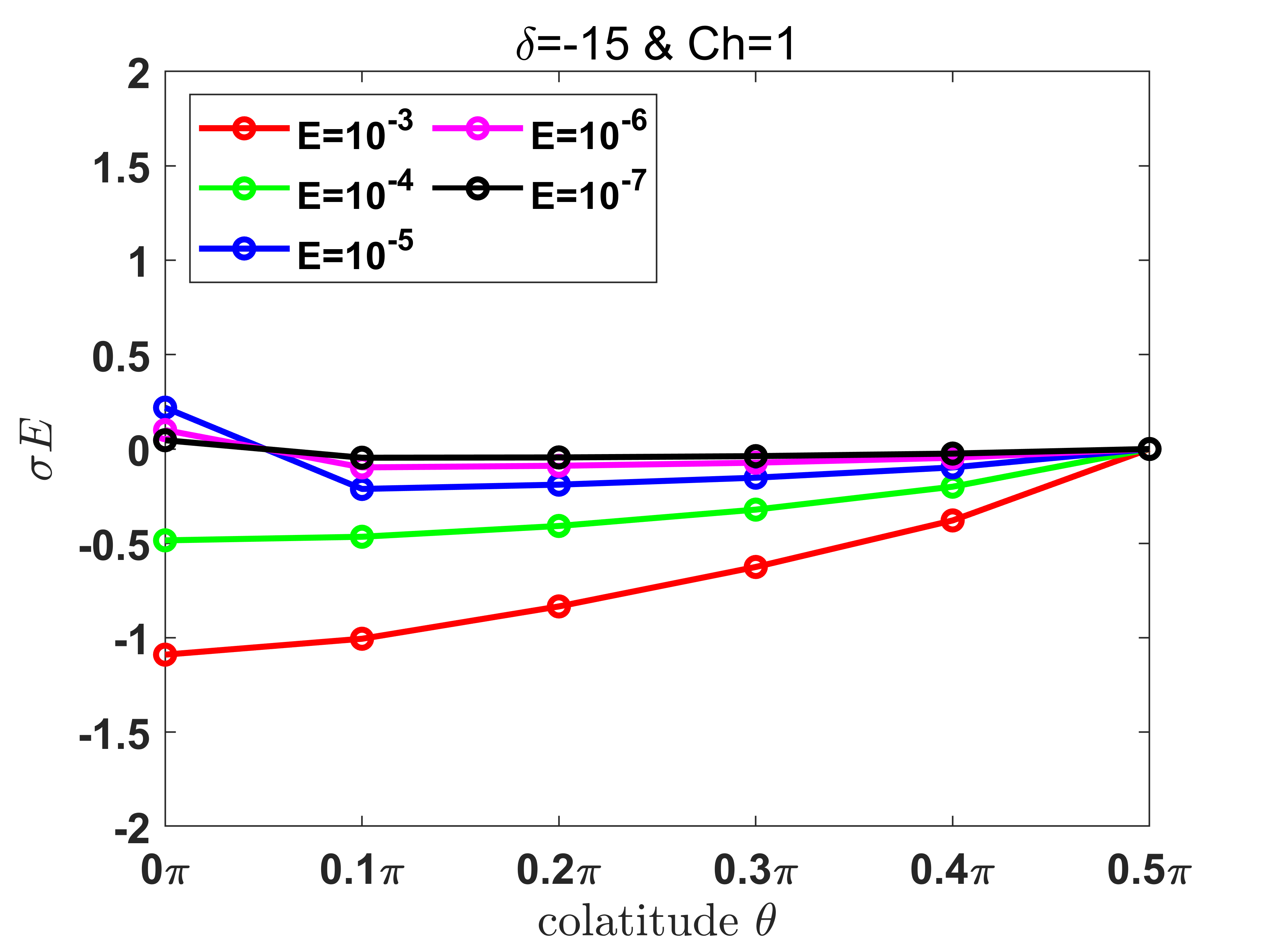}}
    \subfigure[]{
    \includegraphics[width=0.45\columnwidth]{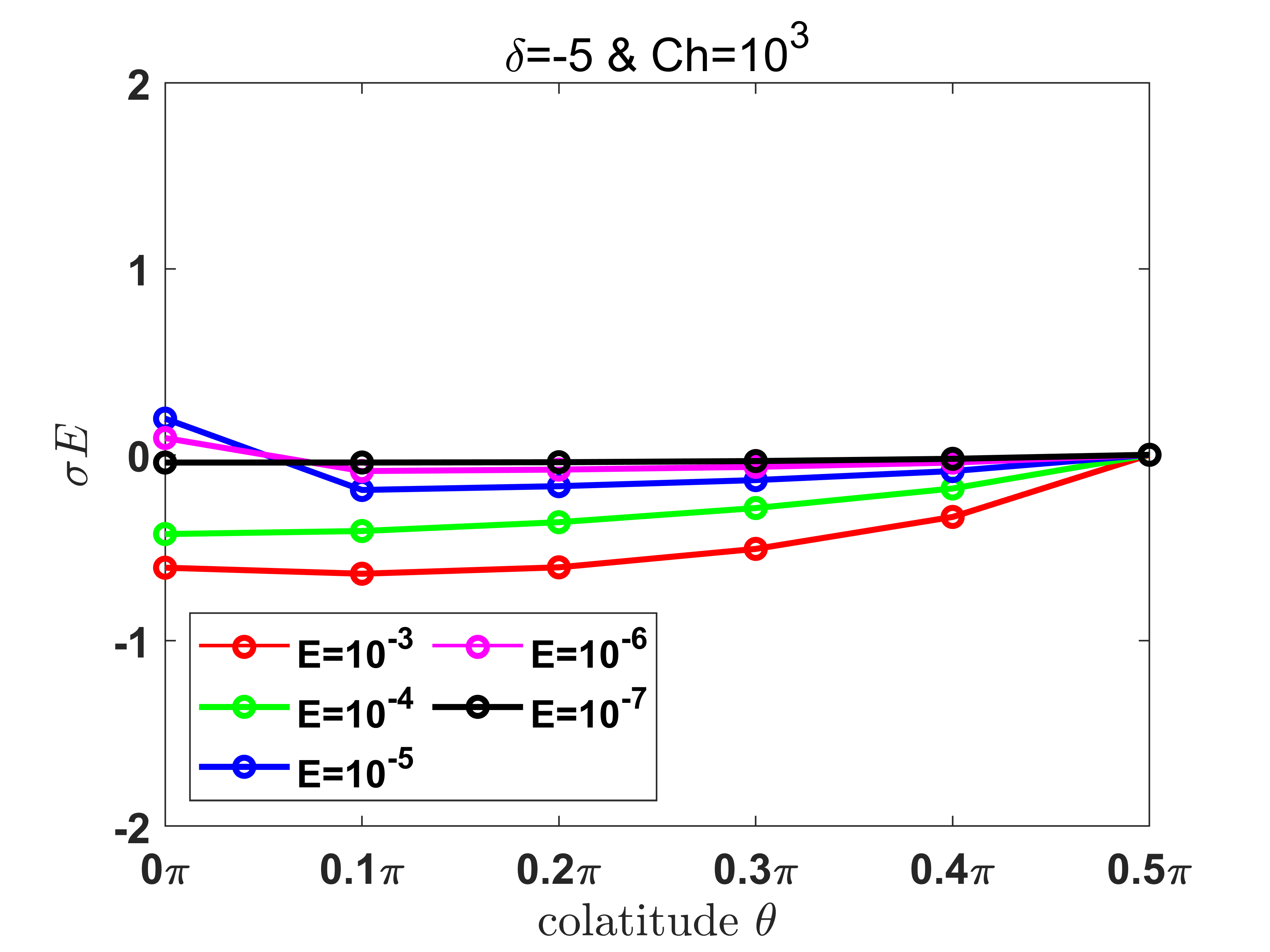}}
    \subfigure[]{
    \includegraphics[width=0.45\columnwidth]{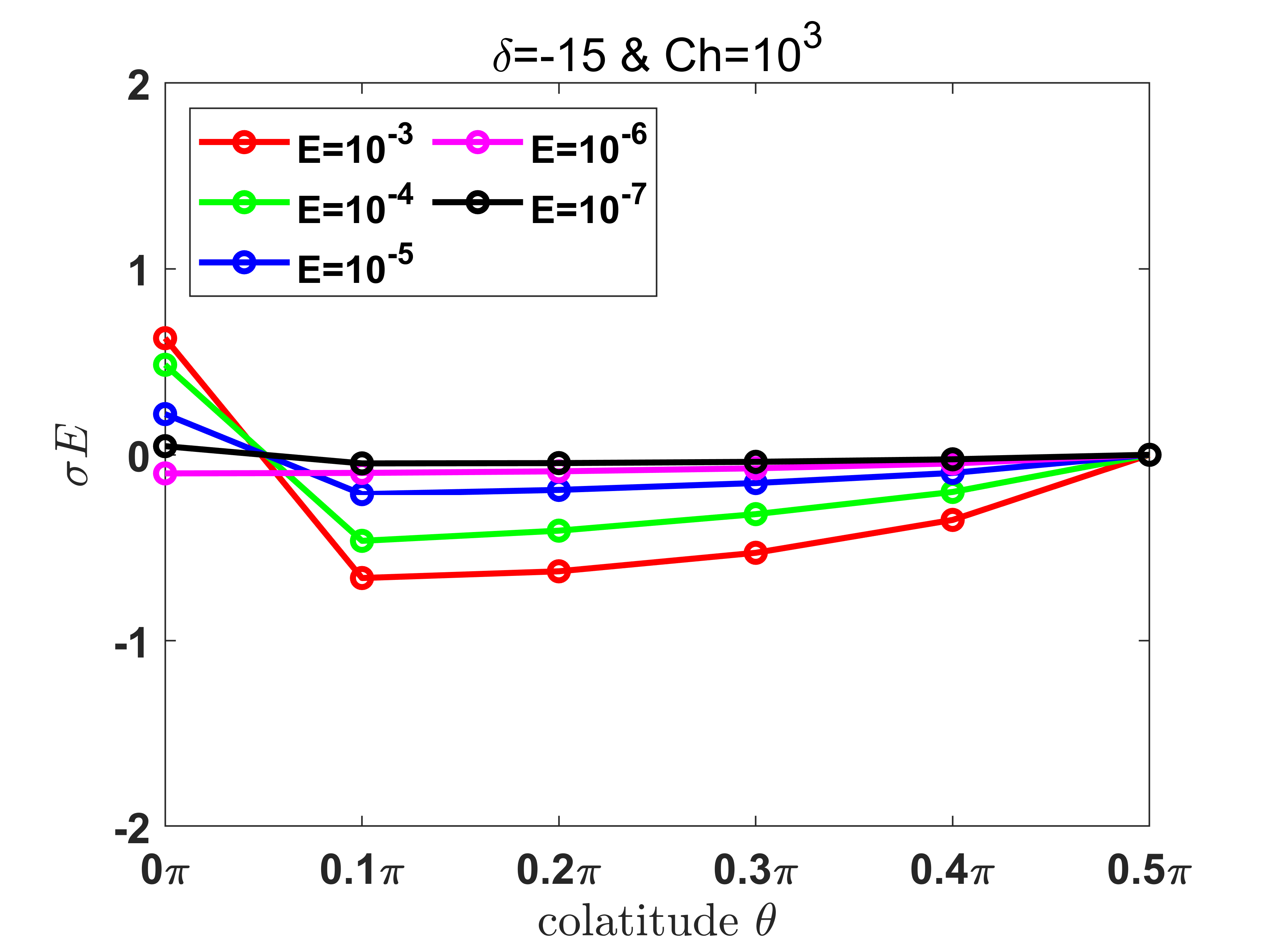}}
    \subfigure[]{
    \includegraphics[width=0.45\columnwidth]{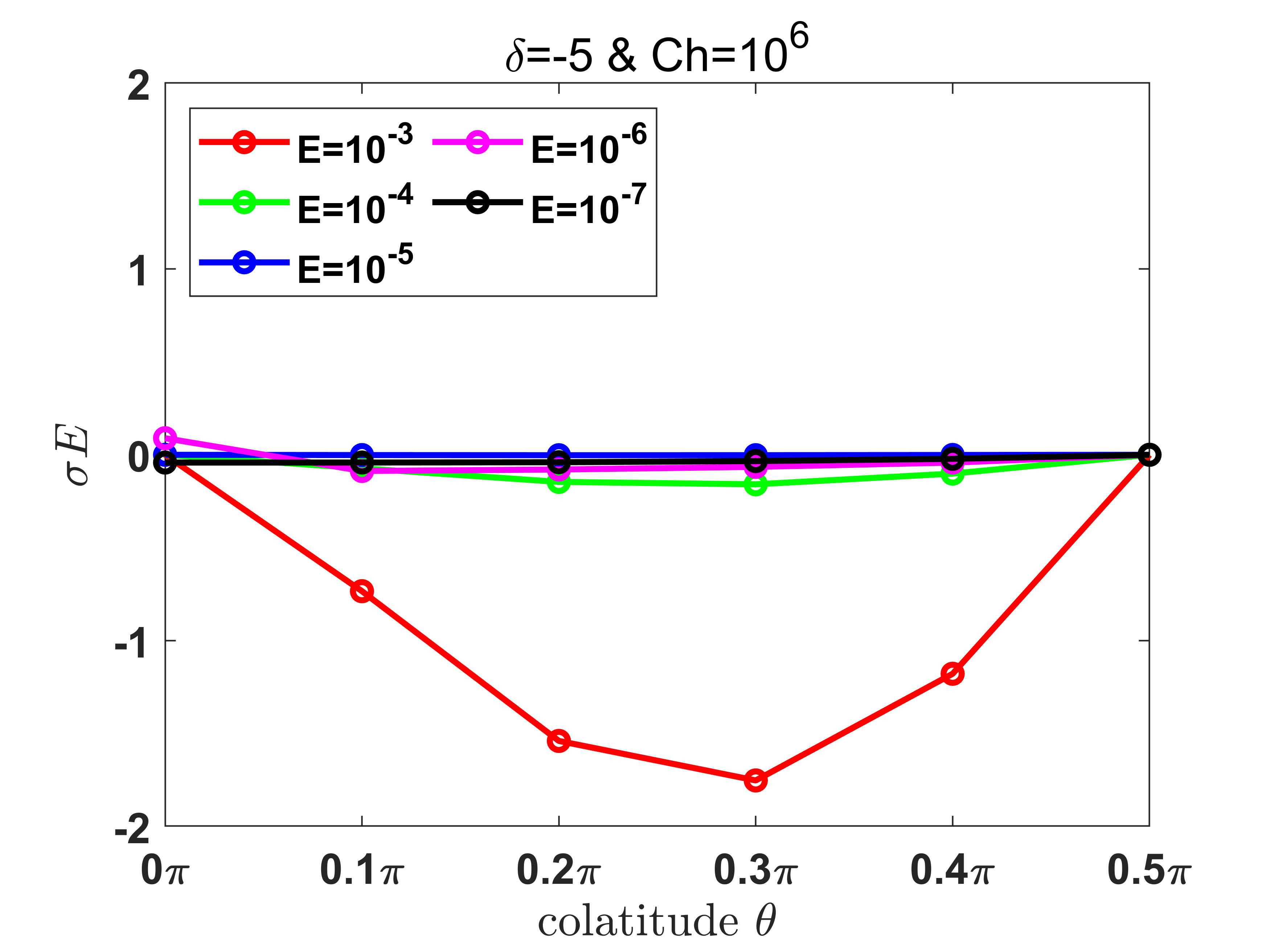}}
    \subfigure[]{
    \includegraphics[width=0.45\columnwidth]{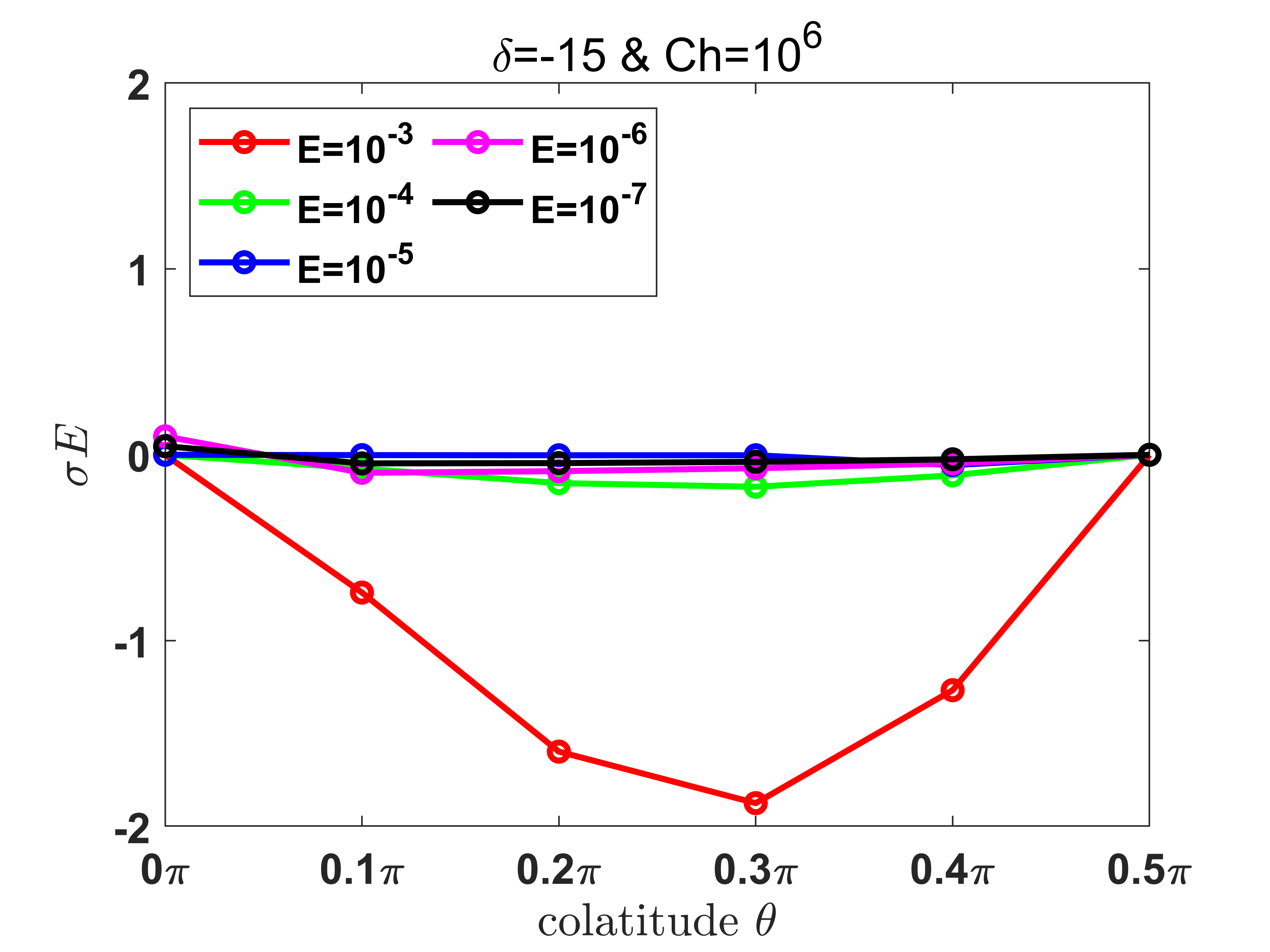}}
    \caption{Wave frequencies $\sigma E$ for cases $\delta=-5$ or $\delta=-15$ when $\mathbf{B}_{0}$ is parallel to $\mathbf{\Omega}$. Panels (a)-(b), (c)-(d), (e)-(f) show the results of $Ch=1$, $Ch=10^{3}$ and $Ch=10^{6}$, respectively.}
    \label{fig:f4}
\end{figure}

To further elucidate the influence of the background magnetic field on rotating convection, we present $\sigma E$ for different cases in Fig.~\ref{fig:f4}. In pure rotating convection, the frequency of inertial waves must satisfy the condition $|\sigma| \leq \sigma_{I}=2E^{-1}$. Conversely, in pure magneto convection, the Alfv\'en wave frequency is given by $\sigma_{A}=(Pm Ch^{-1})^{1/2}$. In convection involving both rotation and a background magnetic field, the inertial waves undergo modification by the background magnetic field, resulting in magneto-inertial waves. These two frequencies can be compared using $\sigma_{A}/\sigma_{I}=2(Pm^{-1} Ch E^{2})^{1/2}$. For our rapid rotation calculations with an Ekman number of $E\leq 1\times 10^{-3}$, the ratio $\sigma_{A}/\sigma_{I} \ll 1$ at $Ch=1$ and $Ch=10^3$.  In these cases, the magneto-inertial waves are primarily influenced by the inertial effect. However, for $Ch=10^{6}$, the ratio $\sigma_{A}/\sigma_{I}$ can exceed one for $E=10^{-3}$. This implies that a strong magnetic field can significantly alter inertial waves. As depicted in Fig.~\ref{fig:f4}, all wave frequencies satisfy the condition, suggesting that the waves behave as inertial waves. We also observe a general decrease in $\sigma E$ as $Ch$ increases, with the value of $\sigma E$ nearing zero for small $E$ at $Ch=10^6$.  This indicates that the magnetic field tends to suppress the inertial waves. A special case arises at $E=10^{-3}$ and $Ch=10^{6}$. In this case, the wave frequency experiences a substantial increase in the mid-latitude. As mentioned above, $\sigma_{A}$ is greater than $\sigma_{I}$ in this case, making it possible for the magneto-inertial waves dominated by the magnetic effect.

\begin{figure}
    \centering
    \subfigure[]{
    \includegraphics[width=0.45\columnwidth]{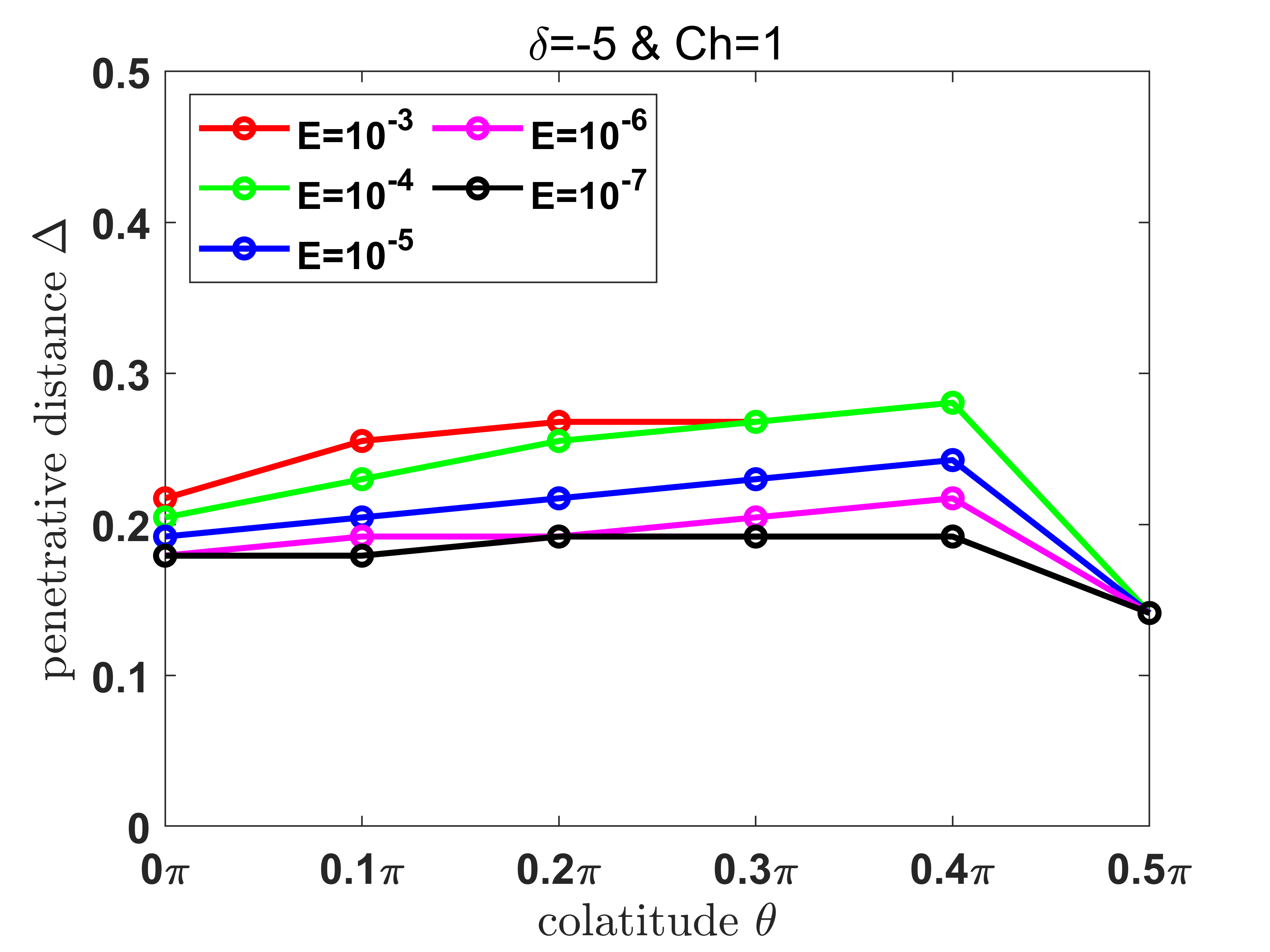}}
    \subfigure[]{
    \includegraphics[width=0.45\columnwidth]{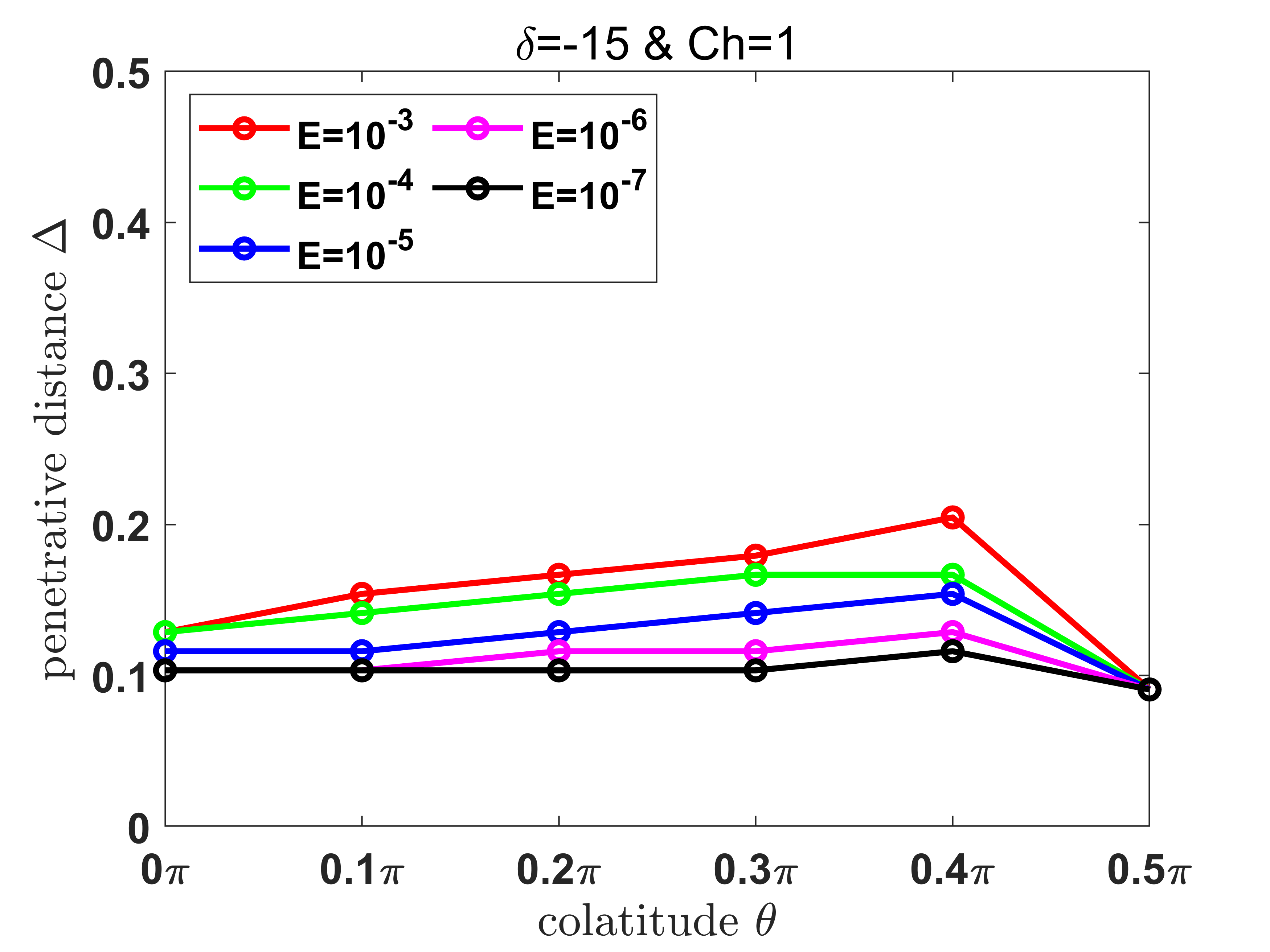}}
    \subfigure[]{
    \includegraphics[width=0.45\columnwidth]{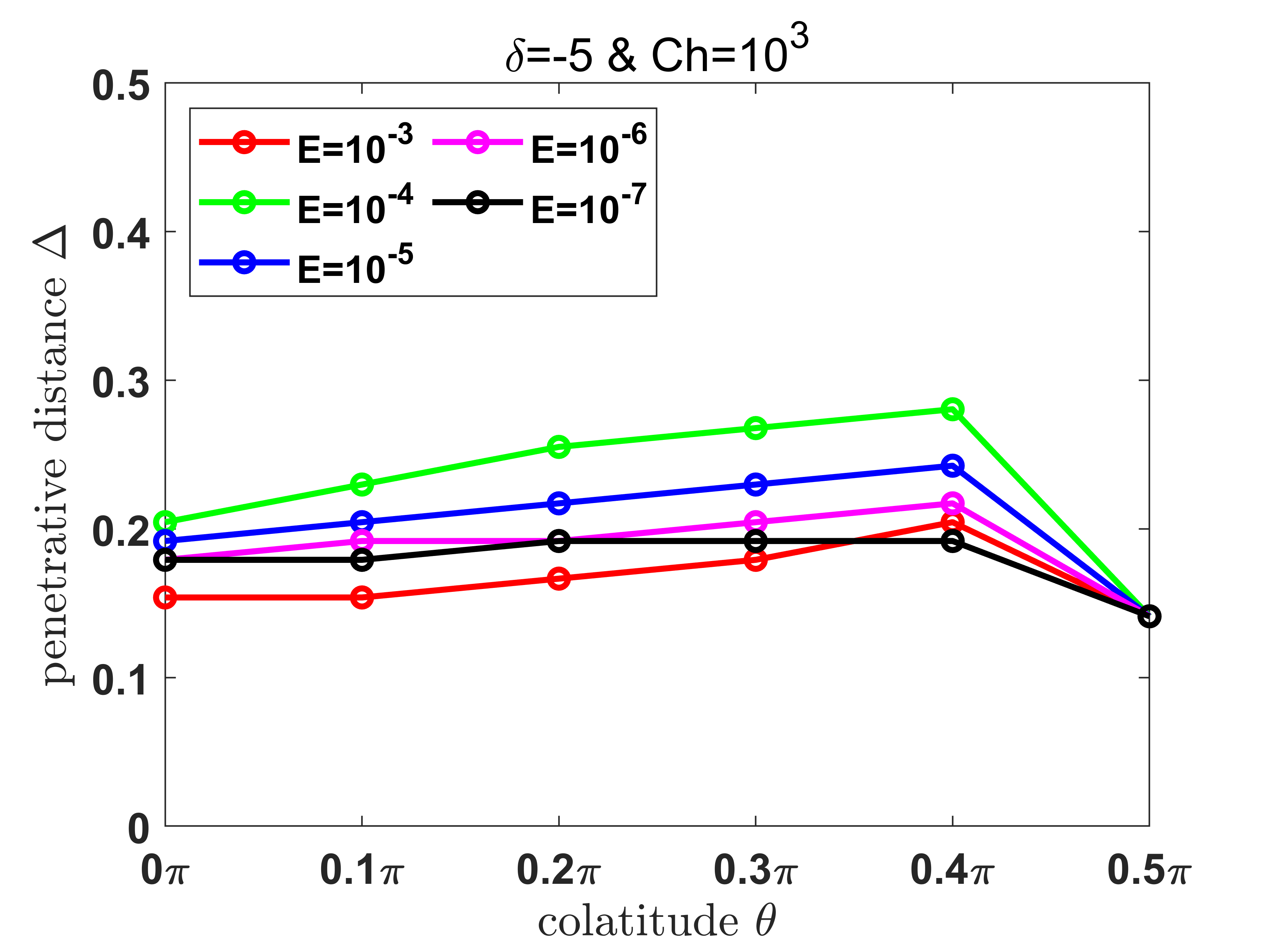}}
    \subfigure[]{
    \includegraphics[width=0.45\columnwidth]{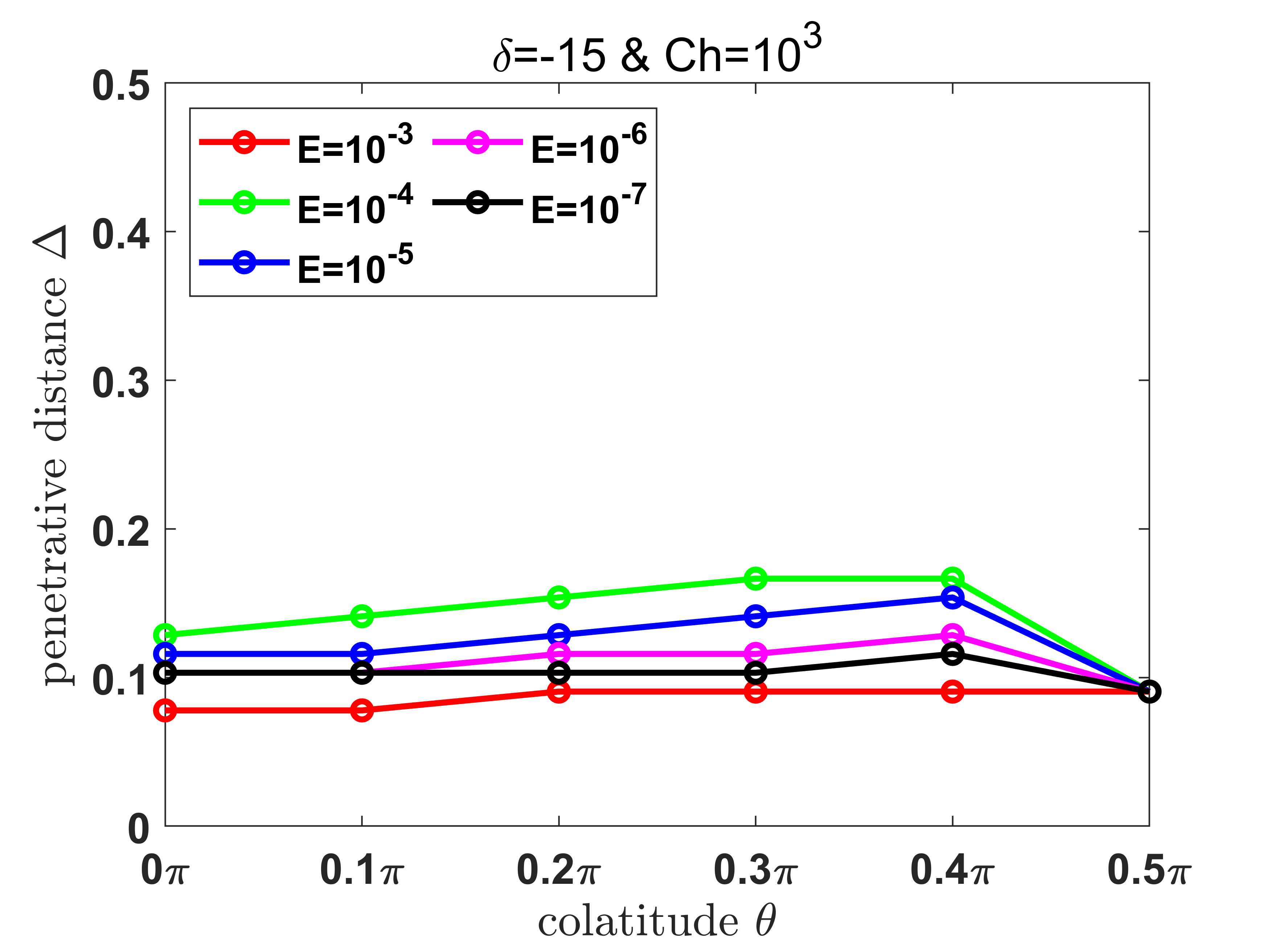}}
    \subfigure[]{
    \includegraphics[width=0.45\columnwidth]{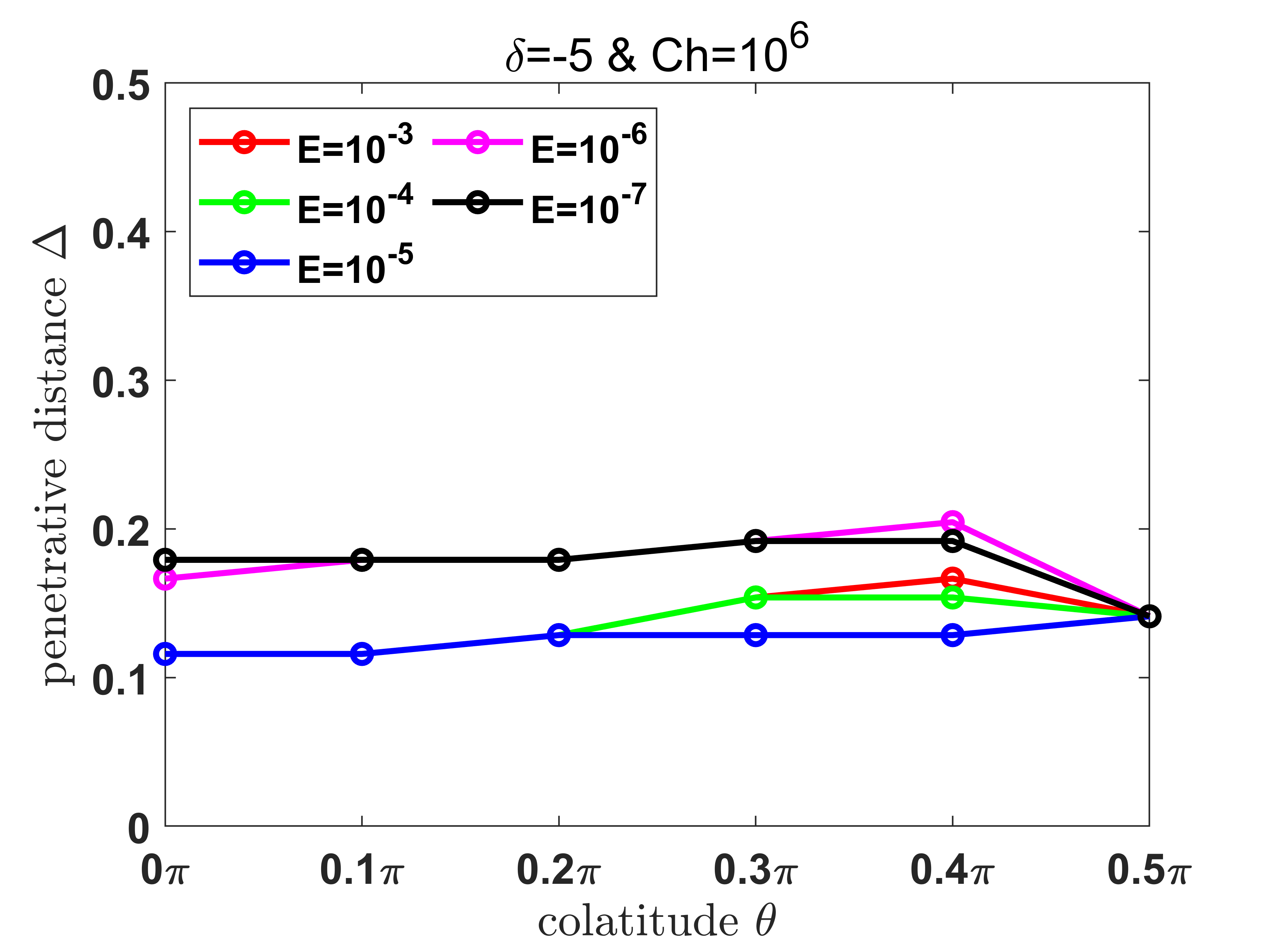}}
    \subfigure[]{
    \includegraphics[width=0.45\columnwidth]{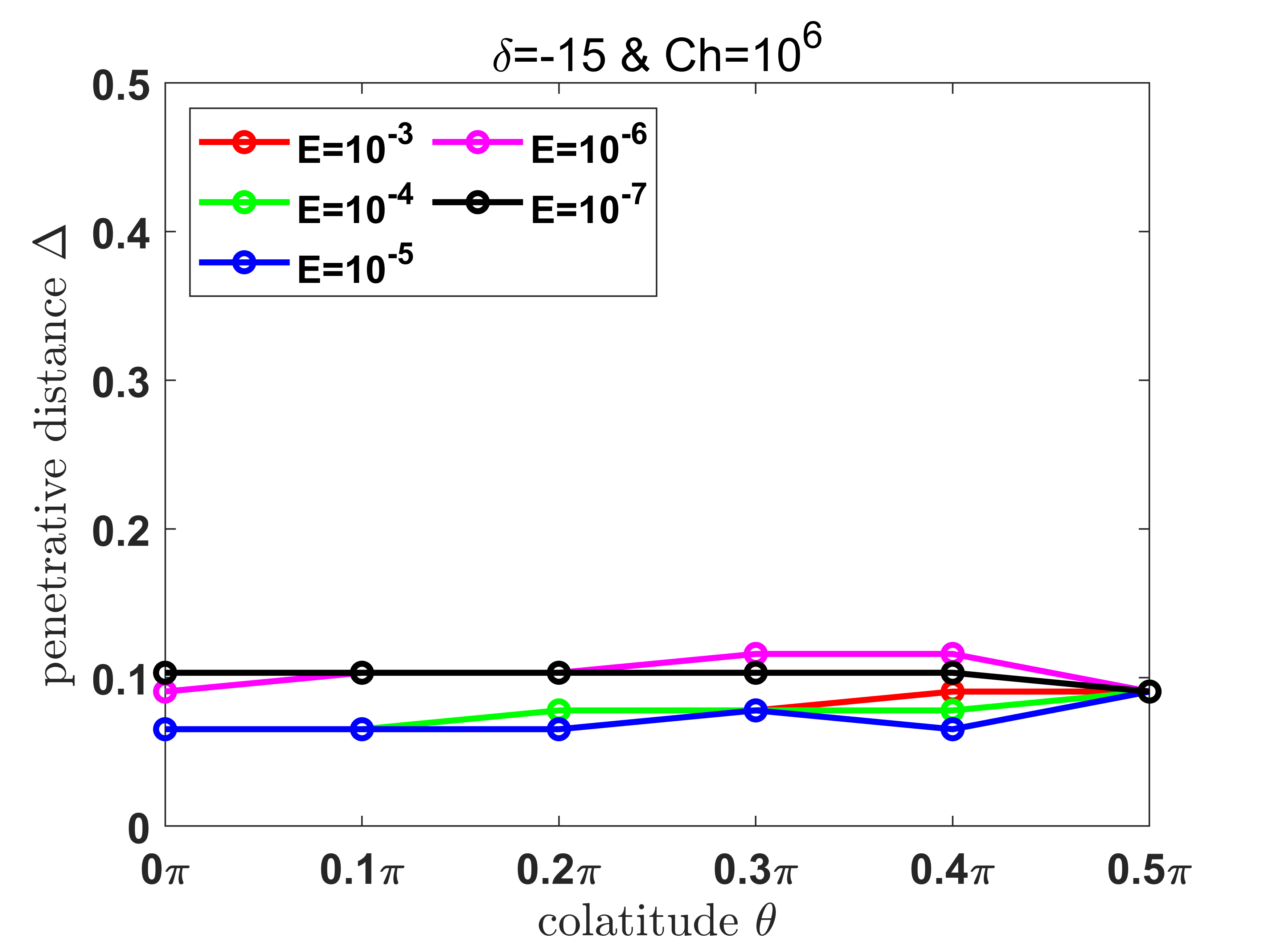}}
    \caption{Companion to Fig.~\ref{fig:f3} but with the interface located at $z=1/3$.}
    \label{fig:f5}
\end{figure}

In our computations, we primarily concentrate on configurations where the interface is situated at the midplane $z=0.5$. Additionally, we have conducted a separate set of simulations with the interface positioned at $z=1/3$ to examine the influence of the interface location. Similar to Fig.~\ref{fig:f3}, which illustrates the result on the interface at $z=0.5$, Fig.~\ref{fig:f5} provides a depiction of the penetrative distances when the interface is positioned at $z=1/3$. It is evident that despite the penetrative distances in Fig.~\ref{fig:f5} being less than those in Fig.~\ref{fig:f3}, the overall patterns exhibit a resemblance. The penetrative distances observed in Fig.~\ref{fig:f5} are approximately $2/3$ of those in Fig.~\ref{fig:f3}. This suggests that changes in the depth of the convective layer have negligible influence on the ratio of penetrative distance to the convective layer. In deep convection of stars, penetrative distances are typically calibrated using the unit of local pressure scale height. The overshooting parameter, defined as the ratio of penetrative distance to the local pressure scale height, is usually set as a constant parameter in stars. However, given the shallow convection zone in our incompressible flow, it is reasonable to calibrate using the unit of the entire convective layer. Given that the depth of the convective layer exerts minimal influence on the ratio, it is sufficient to focus on a single interface at $z=0.5$.

Apart from the interface location, the boundary condition can also influence the outcome. For example, \cite{kolhey2022influence} found that the magnetic boundary conditions can significantly affect the structure of the generated magnetic field. We conducted a comparative analysis in Appendix~\ref{appendixB} by modifying the magnetic boundary condition to be perfectly conducting at the bottom. Our observations indicate that the magnetic boundary condition can significantly impact penetration when $\Lambda >1$. However, when $\Lambda <1$, the effect of the magnetic boundary condition is less pronounced.

\subsubsection{$\mathbf{B}_{0}$ is parallel to $\mathbf{\hat{z}}$}
\begin{figure}
    \centering
    \subfigure[]{
    \includegraphics[width=0.45\columnwidth]{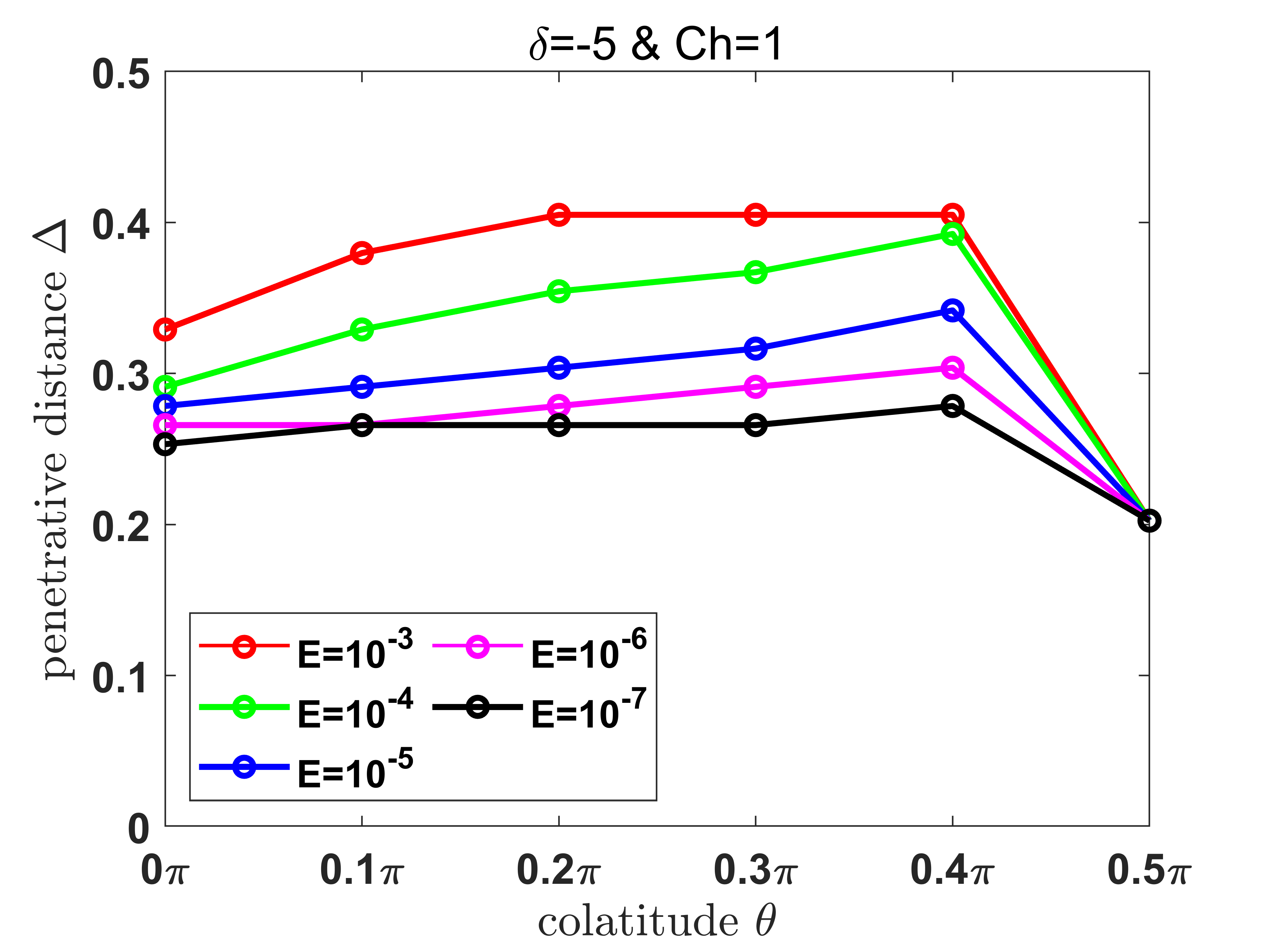}}
    \subfigure[]{
    \includegraphics[width=0.45\columnwidth]{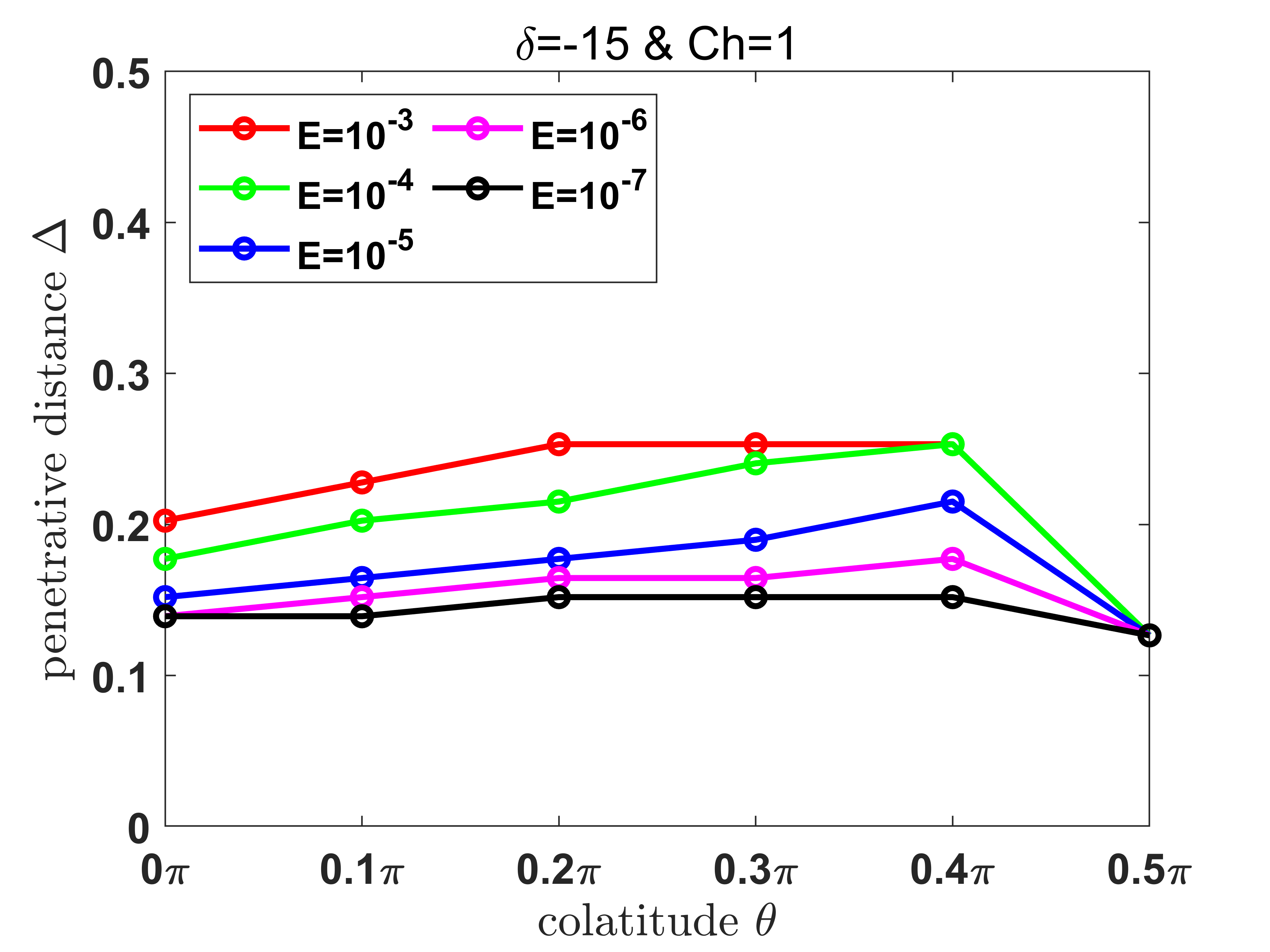}}
    \subfigure[]{
    \includegraphics[width=0.45\columnwidth]{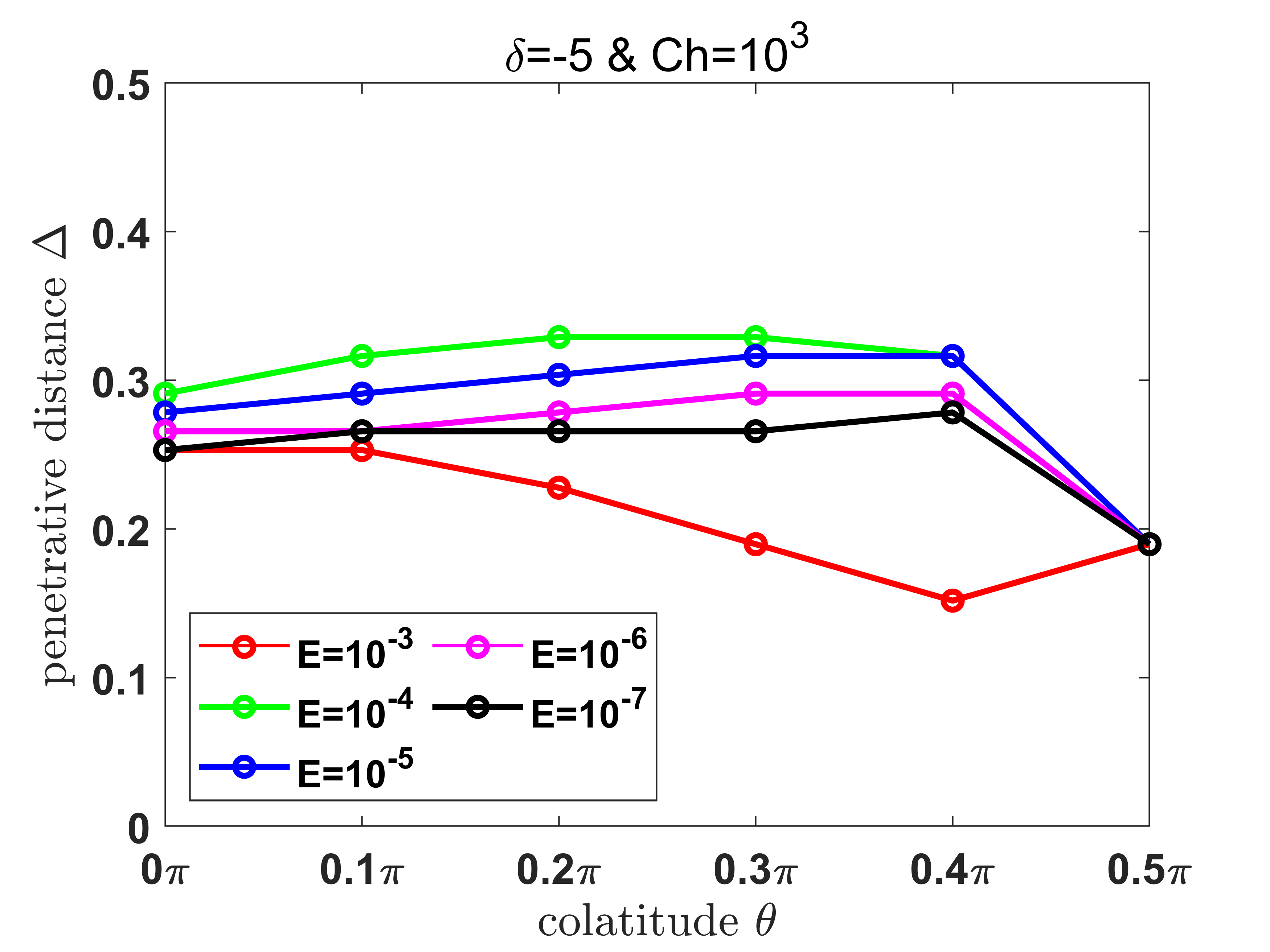}}
    \subfigure[]{
    \includegraphics[width=0.45\columnwidth]{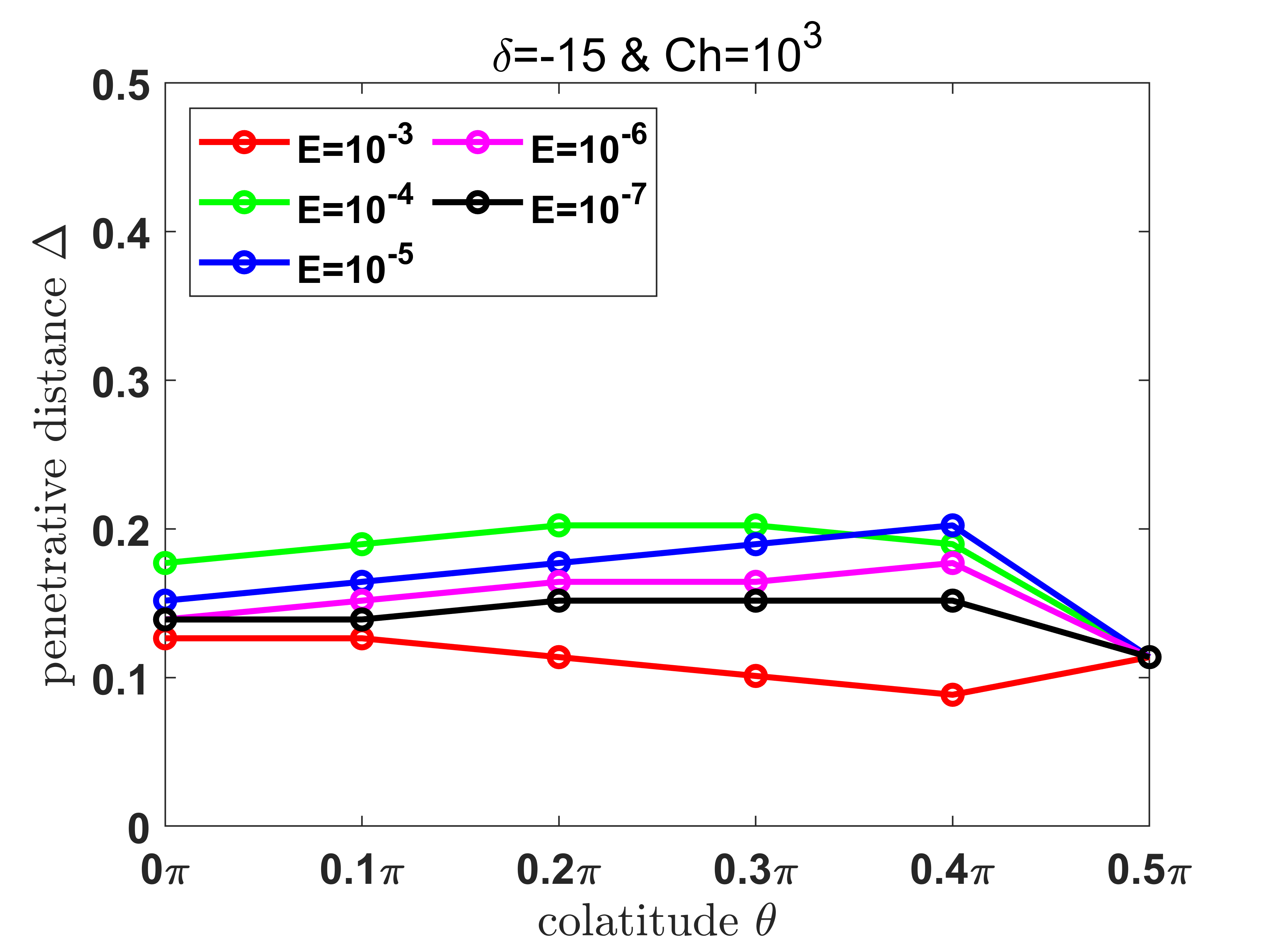}}
    \subfigure[]{
    \includegraphics[width=0.45\columnwidth]{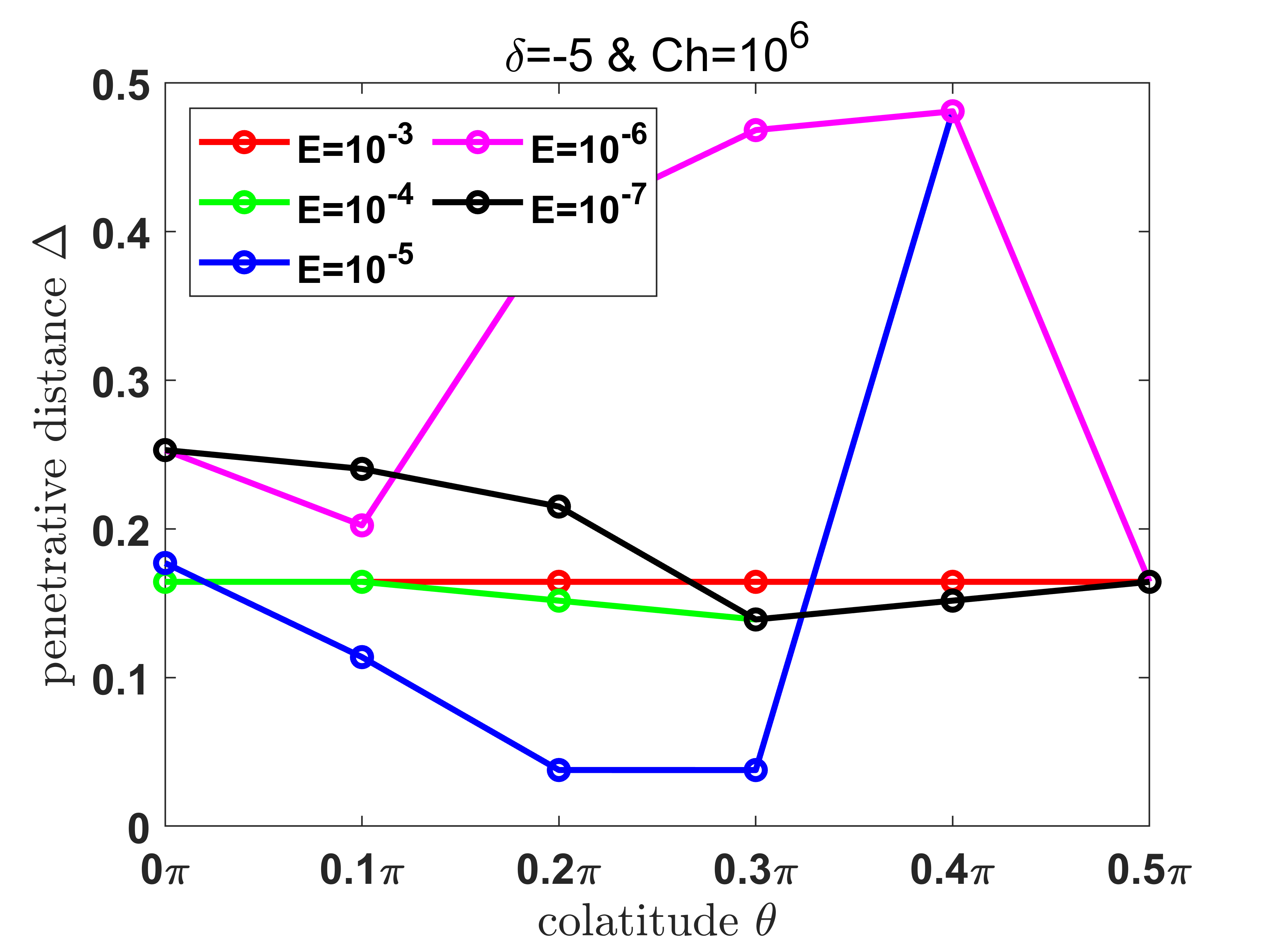}}
    \subfigure[]{
    \includegraphics[width=0.45\columnwidth]{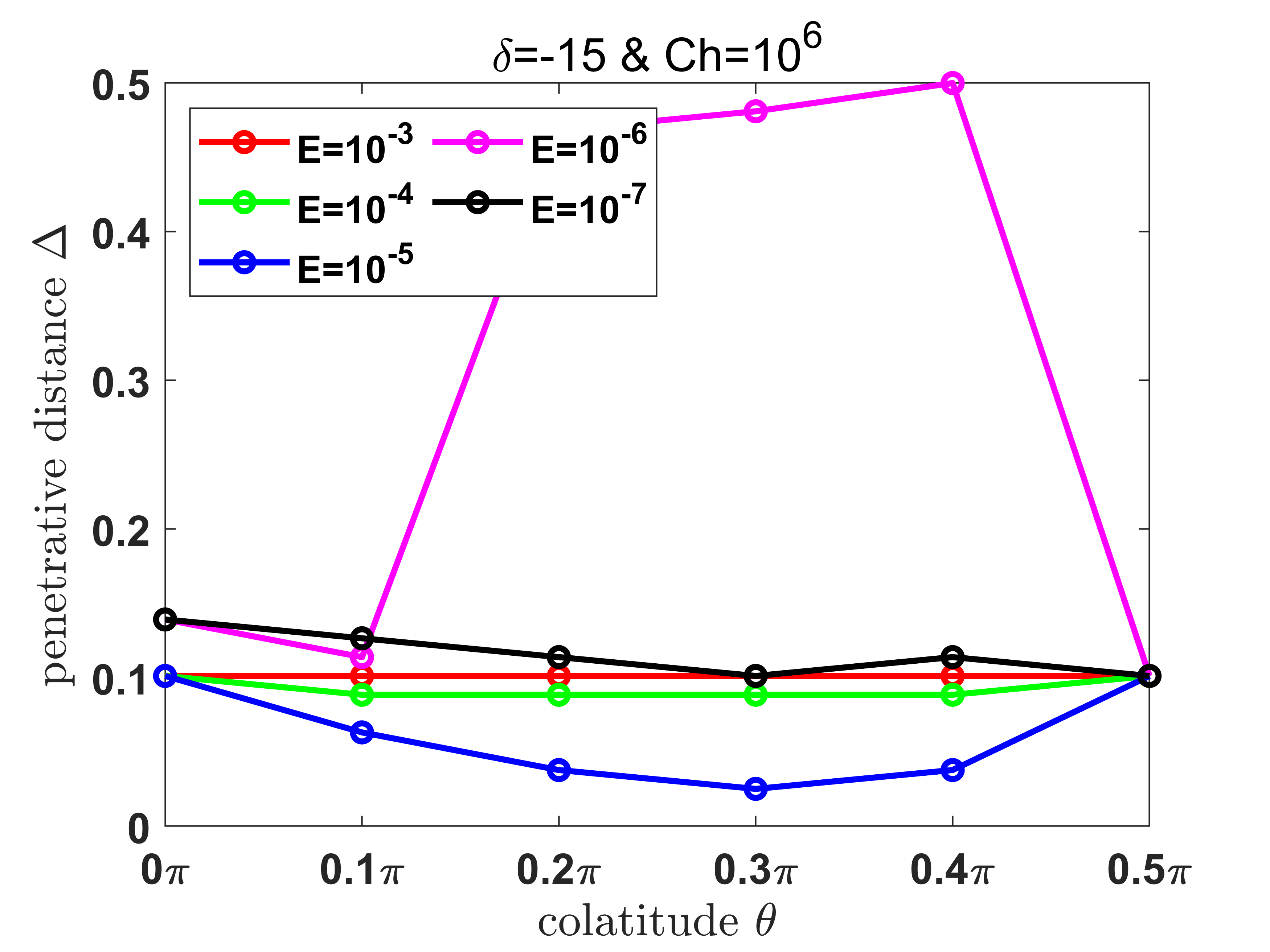}}
    \caption{Penetrative distances for cases $\delta=-5$ or $\delta=-15$ when $\mathbf{B}_{0}$ is parallel to $\mathbf{\hat{z}}$. Panels (a)-(b), (c)-(d), (e)-(f) show the results of $Ch=1$, $Ch=10^{3}$ and $Ch=10^{6}$, respectively.}
    \label{fig:f6}
\end{figure}

\begin{figure}
    \centering
    \subfigure[]{
    \includegraphics[width=0.45\columnwidth]{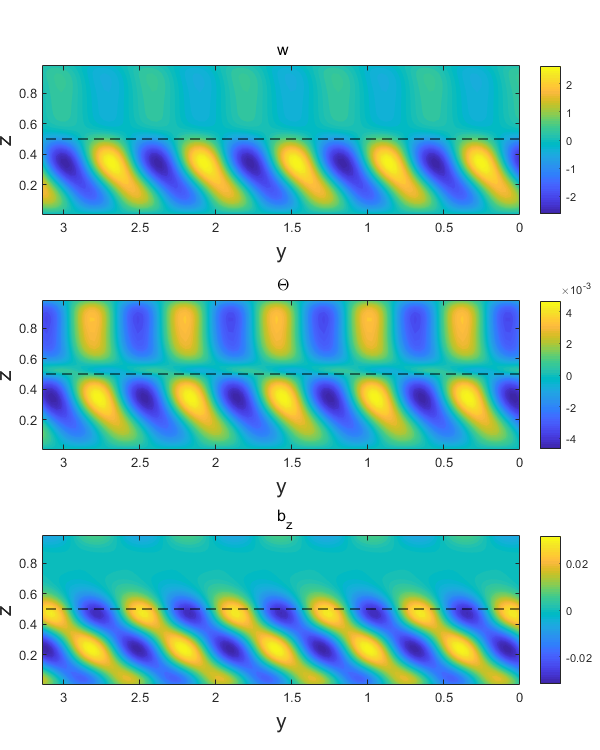}}
    \subfigure[]{
    \includegraphics[width=0.45\columnwidth]{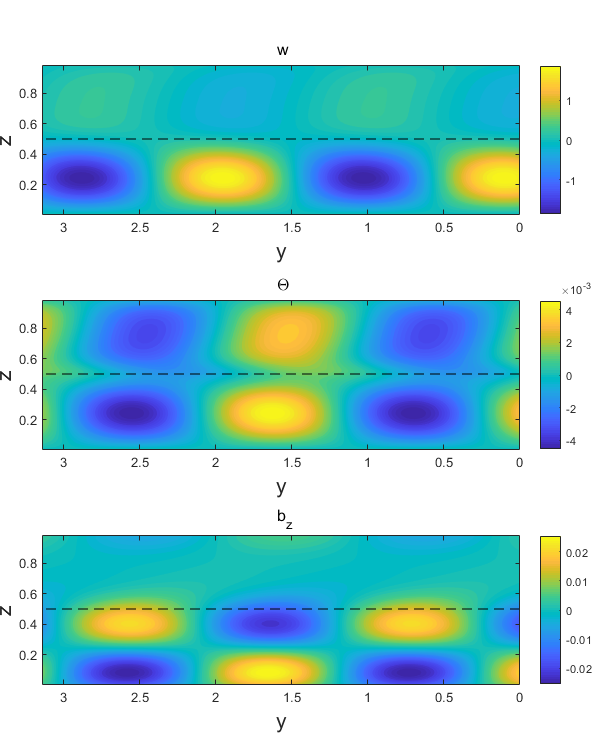}}
    \subfigure[]{
    \includegraphics[width=0.45\columnwidth]{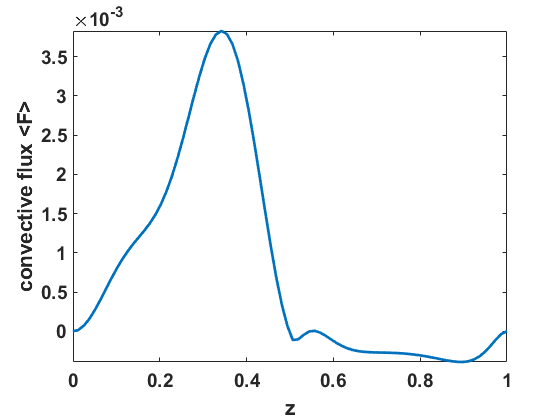}}
    \subfigure[]{
    \includegraphics[width=0.45\columnwidth]{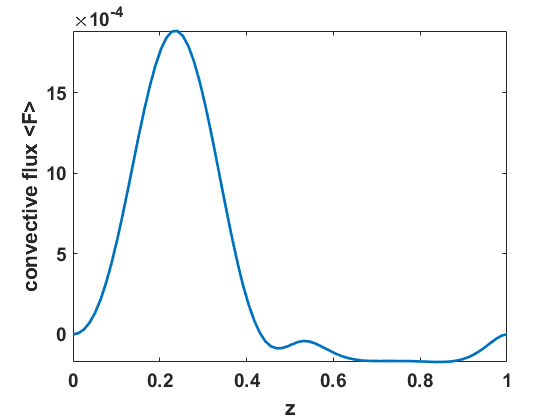}}
    \caption{The contour plots for the vertical velocity $w$, the temperature perturbation $\Theta$, and the vertical component of magnetic perturbation $b_{z}$, at $\delta=-5$, $\theta=0.4\pi$, $Pr=Pm=0.1$, $E=10^{-3}$, $Ch=10^{6}$. The left and right columns use different $\gamma$ with (a)$\gamma=\theta$ and (b)$\gamma=0$, respectively. (c) and (d)The corresponding convective flux $\langle F\rangle$ as a function of $z$.}
    \label{fig:f7}
\end{figure}

When $\mathbf{B}_{0}$ is parallel to $\mathbf{\hat{z}}$, it results in $\gamma=0$. We have conducted a comparative study with those of $\gamma=\theta$. Fig.~\ref{fig:f6} illustrated the corresponding penetrative distances when $\gamma=0$. It is clear that the penetrative distances do not significantly deviate from those of $\gamma=\theta$ when the magnetic effect is weak, i.e., $\Lambda \ll 1$ (Figs.~\ref{fig:f6}(a-b)). However, in a flow dominated by magnetism where $\Lambda \gg 1$. we observe notable differences. Abrupt changes in penetrative distances occur when $E$ decreases from $10^{-5}$ to $10^{-7}$, as seen in Figs.~\ref{fig:f6}(e-f). To understand this, we plotted in Fig.~\ref{fig:f7} the contours of $w$, $\Theta$, and $b_{z}$ for the cases $\theta=0.2\pi$ and $\theta=0.4\pi$ with $E=10^{-5}$ of Fig.~\ref{fig:f6}(e). Interestingly, we found that the contours of $w$ and $\Theta$ exhibit different patterns in the unstable and stable layers. In the unstable layer the flow tends to align with the $\mathbf{\Omega}$, while it aligns with the $\mathbf{B_{0}}$ in the stable layer. This suggests that the flow is primarily constrained by fast rotation in the unstable layer and by the strong magnetic field in the stable layer. Additionally, we observe $w$ is weak in the stable layer, while $\Theta$ exhibits strong fluctuations. These strong temperature fluctuations significantly impact penetration. From our computations for the case $\theta=0.2\pi$, we found that the ratio of $|w|^2$ in the stable to unstable layers is about 0.03, while that of $|\Theta|^2$ can reach 0.66. Despite $w$ being small, the penetrative distance can be large if $\langle F\rangle$ is used as a proxy. This poses a challenging in quantifying the penetrative distance. In both cases, the structures of $\langle F\rangle$ are similar. However, for the case $\theta=0.2\pi$, $\langle F\rangle$ crosses zero in the unstable layer, while for the case $\theta=0.4\pi$, it does not. We define the penetrative distance as the point where the convective flux has attained 5\% of its most negative value. As a result, the calculated penetrative distance is small for $\theta=0.2\pi$, but large for $\theta=0.4\pi$. However, under such circumstances, the penetrations of momentum and energy are likely to differ. Consequently, solely using $\langle F\rangle$ as the proxy for defining the penetrative distance may not be suitable. An accurate assessment of the penetrative distance likely requires considering the combined effects of the self- and cross-correlations of $w$ and $\Theta$. Conducting three-dimensional numerical simulations of penetrative convection with passive scalars could potentially enhance our understanding of how to select appropriate proxies, which is beyond the scope of the current research.

\section{Summary}
This paper explores the penetrative convection in the presence of a background magnetic field within rotating, tilted $f$-plane by linear stability analysis. We integrate wave theory and convection theory to explain the penetrative convection in the rotating magneto flow. Various wave types, including inertial waves, internal gravity waves, and slow magnetic waves, can be generated in the penetrative convection of a rotating magneto flow. When these waves transmit through the interface between convectively unstable and stable layers, and if wave solutions are permitted in both sides, resonance may occur if the wavenumbers on either side are comparable. This resonance enables efficient penetration from the unstable layer to the stable one. The resonance usually occurs in flow with weakly stratified stable layer at critical latitudes. The flow can penetrate through all the stable layer when the resonance takes place. What's more, when the intensity of the stable layer approaches to zero $\delta \rightarrow 0$, the flow motions in the stable layer can be more vigorous than those in the unstable layer, where so called `teleconvection' occurs. However, in a strongly stratified stable layer, the wave becomes evanescent. The penetrative distance is then determined by the decay rate of the evanescent wave.

In our study, we computed the decay rate for a rapidly rotating magneto-convective flow, characterized by strongly stratified stable layers, within the context of tilted $f$-planes.
The relative strength of magnetic to rotational effects can be measured by the Elsasser number $\Lambda$. The flow is dominated by rotation when $\Lambda \ll 1$, and by magnetism when $\Lambda \gg 1$. Our calculations indicate that the penetrative distance in both regimes scales as $O(\delta^{-1/2})$. Consequently, a stronger stratification leads to a reduced penetrative distance. More specifically, in a rotation-dominated flow, we observe a general decrease in penetrative distance with increased rotational effect. Additionally, our computations revealed a minor latitude-dependent increase in penetrative distance within rotation-dominated flow. Interestingly, this increase nearly vanishes as the Ekman number $E$ approaches zero. However, in a magnetism-dominated flow, the penetrative distance generally decreases with increased magnetic effect.

In our calculation, we also observed a significant shift in penetrative distance at $\Lambda \sim 1$. Specifically, when the background field aligns with the rotational axis, we found that in the absence of a magnetic field, the penetrative distance typically decreases with the rotational effect. However, in the presence of a magnetic field, the penetrative distance tends to decrease when $\Lambda<1$ and increase when $\Lambda>1$ with the rotational effect. This is attributed to the transition of the flow state from rotation-dominated to magnetism-dominated around $\Lambda\sim 1$. Our calculation underscores the significance of $\Lambda$ when examining penetration in rotating magneto-convection.

We have also examined the penetration when the background magnetic field does not align with the rotational axis. The penetrative distance remains nearly identical to the case where $\mathbf{B_{0}}$ is parallel to $\mathbf{\Omega}$, given  $\Lambda>1$. However, we observe notable differences when $\Lambda<1$. Firstly, the flow exhibits distinct patterns in the unstable and stable layers, being constrained by rotation in the unstable layer and by magnetism in the stable layer. Secondly, despite the diminished velocity in the stable layer, the temperature fluctuation can be substantial. Thirdly, the penetrative distance measured by indicators of momentum and energy can be significantly different. Further investigation is required to define the penetrative distance appropriately.

In this study, we employ a linear stability analysis of the onset of convection to investigate the penetration in rotating magneto-convection. However, the conditions in stars and planets can be significantly different. In actual stars or planets, the Rayleigh number can far surpass its critical value for the onset of convection, leading to highly turbulent flows. As evidenced in simulations of turbulent convection \citep{korre2019convective,cai2020aupward}, turbulent mixing can play a crucial role in penetrative convection, which has not been taken into account in our linear analysis. Furthermore, the emergence of large-scale vortices in rapidly rotating convection could probably alter the penetrative distance substantially \citep{chan2007rotating,kapyla2011starspots,rubio2014upscale,guervilly2014large,cai2021large}. Despite its simplicity, the linear stability analysis has already demonstrated significant differences in penetration behaviors when $\Lambda\ll 1$ and $\Lambda\gg 1$. Conducting simulations of rapidly rotating convection with a background magnetic field within a tilted $f$-plane could further enrich our understanding in the future.


%
%

%

\begin{acknowledgments}
This work is supported by the National Natural Science Foundation of China (No. 12173105), and Guangdong Basic and Applied Basic Research Foundation (No. 2414050002575) .
\end{acknowledgments}

\section*{Data Availability Statement}
The data that support the findings of this study are available from the corresponding author upon reasonable request.

\appendix
\section{Example on the calculation of the penetrative distance}\label{appendixA}
\begin{figure}
    \centering
    \includegraphics[width=0.45\columnwidth]{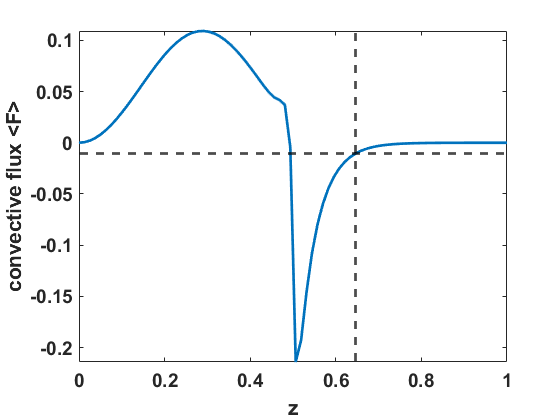}
    \caption{Convective flux as a function of height for the case at $Pr=Pm=0.1$, $E=1\times 10^{-6}$, $Ch=1$, and $\theta=\gamma=0$. The cross between the vertical and horizontal dashed lines marks the location ($z=0.646$) where the convection flux has attained 5\% of its most negative value ($-0.214$) in the convectively stable layer.}
    \label{fig:f8}
\end{figure}

Here we show an example on how to quantitatively measure the penetrative distance in our computation. Fig.~\ref{fig:f8} shows the convective flux as a function of height for the case at $Pr=Pm=0.1$, $E=1\times 10^{-6}$, $Ch=1$, and $\theta=\gamma=0$. For this case, the most negative value of the convective flux $-0.214$ is achieved at the location $z=0.506$. The 5\% of the most negative value is $-0.011$, which is achieved at the location $z=0.646$. Since the interface is at the location $z=0.5$, the penetrative distance is $\Delta=0.146$.

\section{Effect of the magnetic boundary condition}\label{appendixB}
In the interiors of stars or planets, the fluid often exhibits high conductivity, making the perfectly conducting magnetic boundary condition a realistic assumption. In this section, we examine the impact of the magnetic boundary condition by designating the lower boundary as perfectly conducting. This is mathematically represented as
\begin{align}
&\tilde{D}_{z}\Psi_{b}(0)=\Psi_{b}(1)=0~,\\
&\tilde{\Phi}_{b}(0)=0~,\\
&D_{z}\tilde{\Phi}_{b}(1)=-a\tilde{\Phi}_{b}(1)~.
\end{align}

\begin{figure}
    \centering
    \subfigure[]{
    \includegraphics[width=0.45\columnwidth]{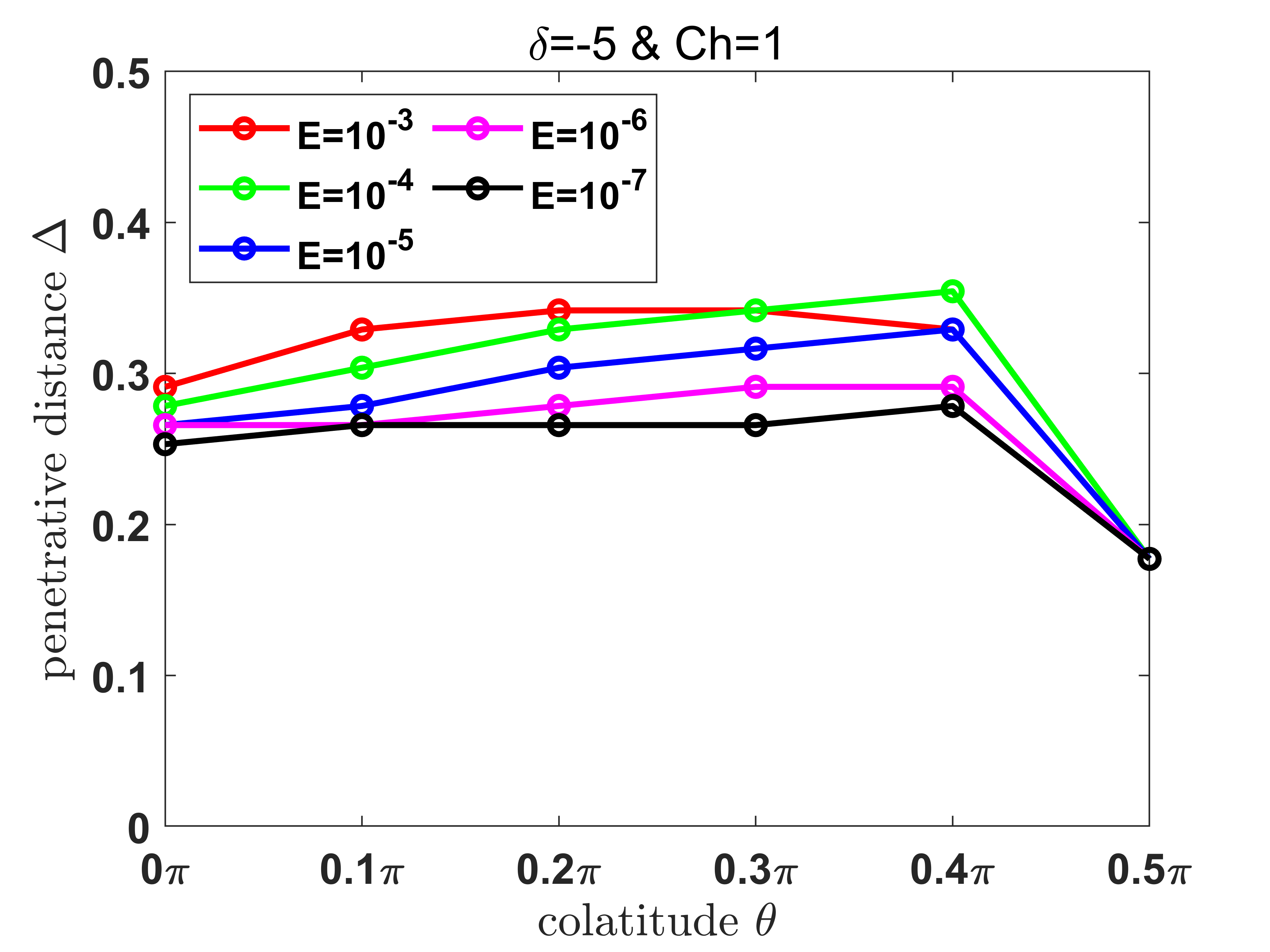}}
    \subfigure[]{
    \includegraphics[width=0.45\columnwidth]{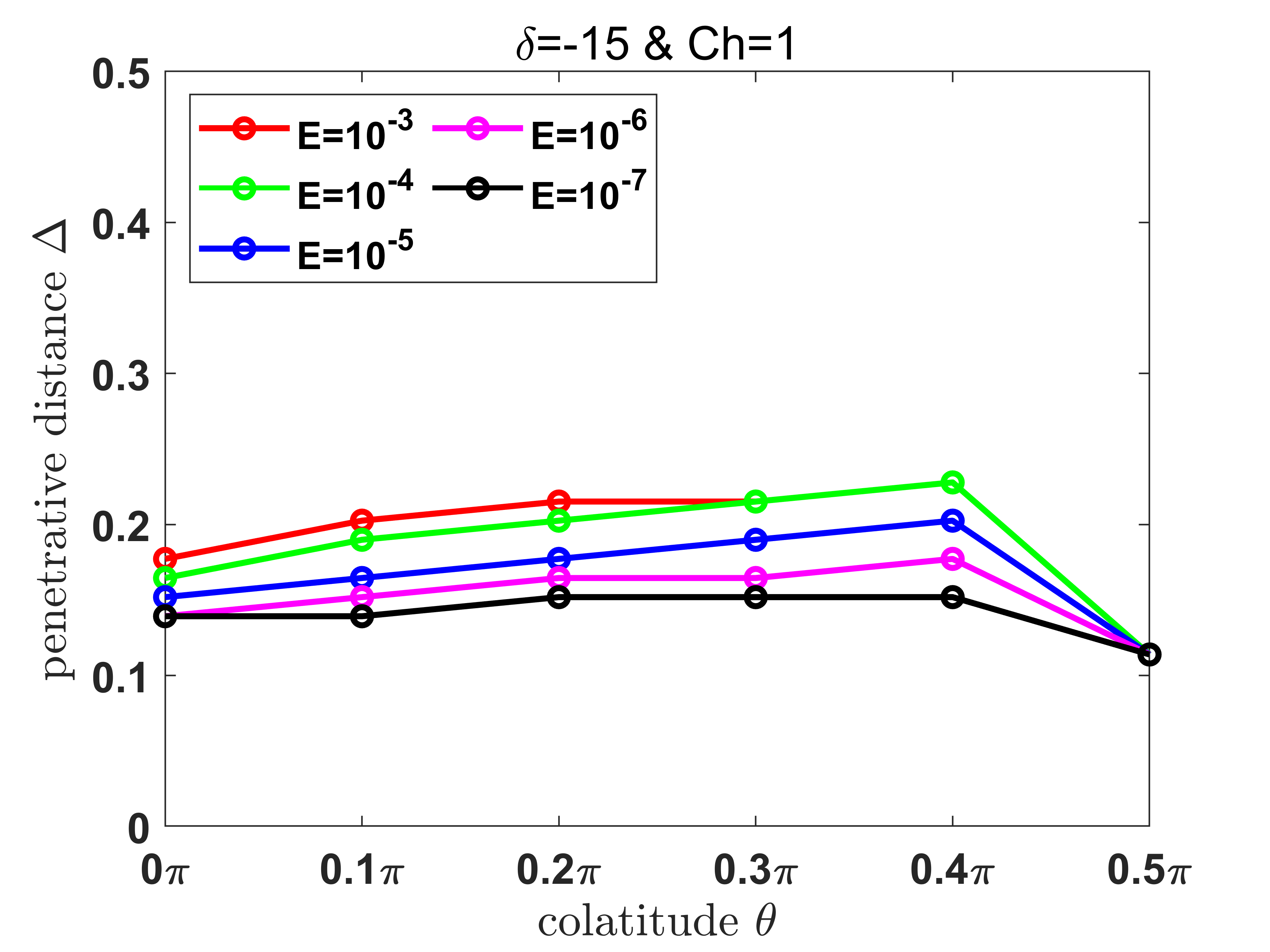}}
    \subfigure[]{
    \includegraphics[width=0.45\columnwidth]{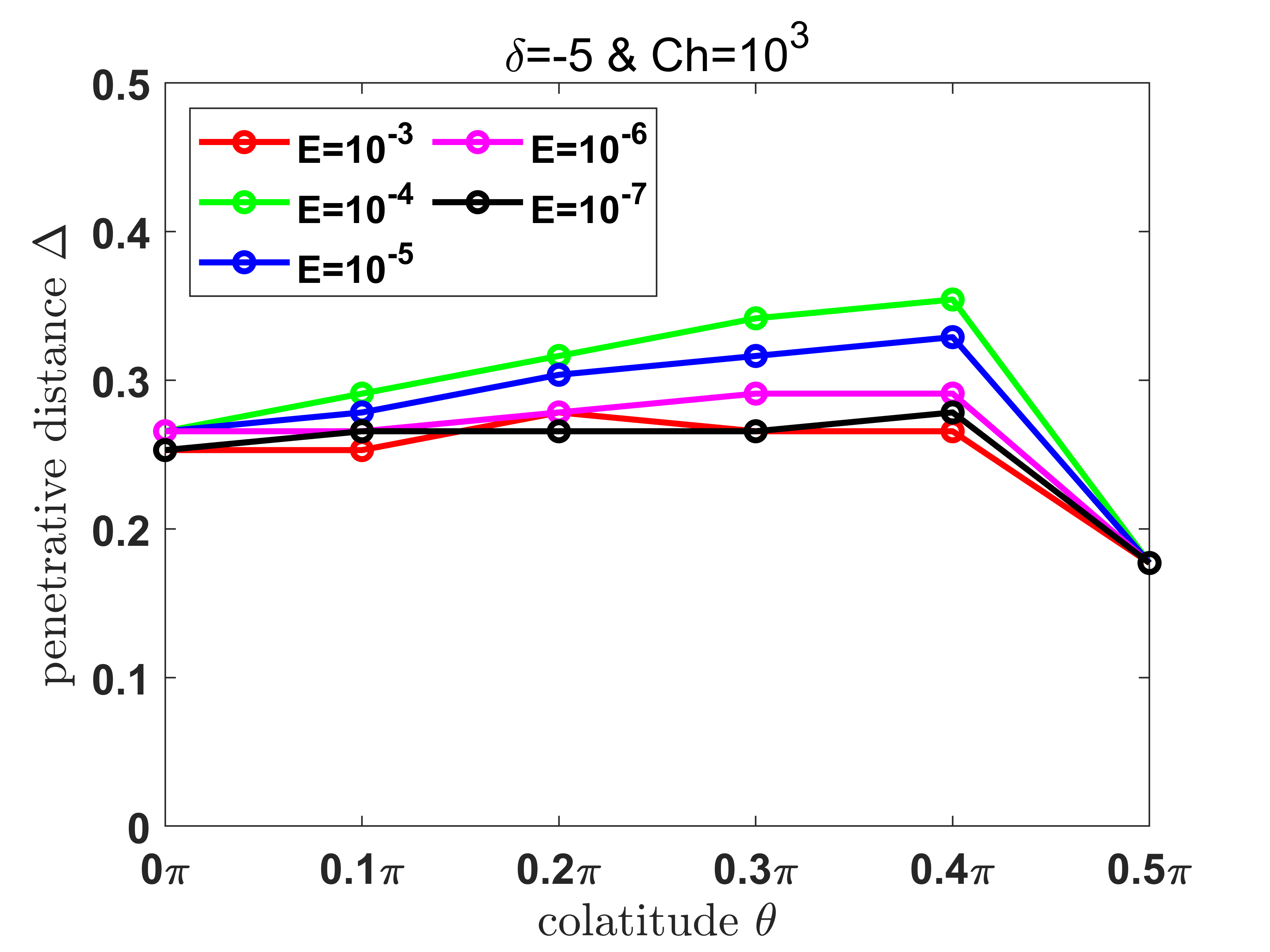}}
    \subfigure[]{
    \includegraphics[width=0.45\columnwidth]{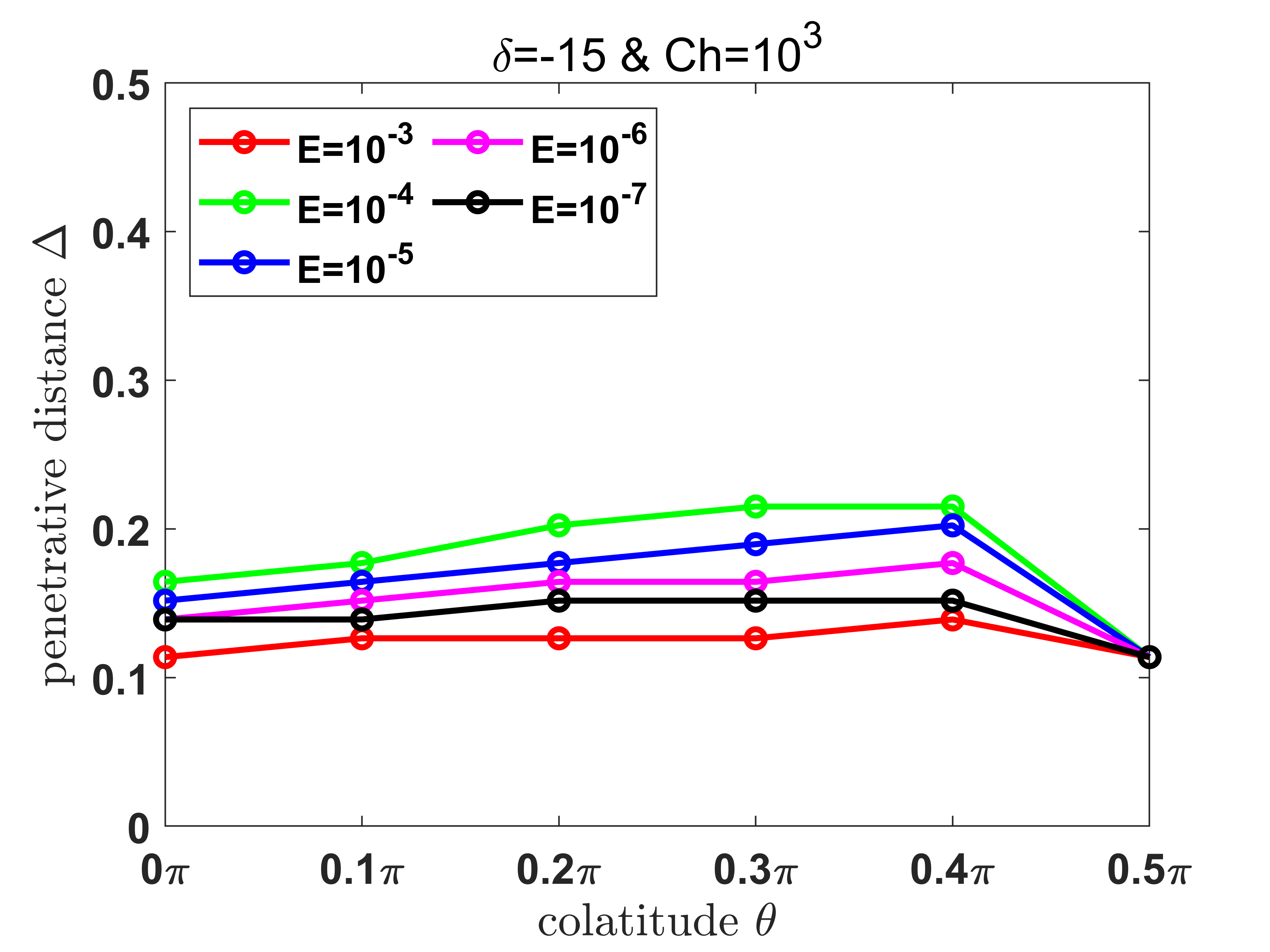}}
    \subfigure[]{
    \includegraphics[width=0.45\columnwidth]{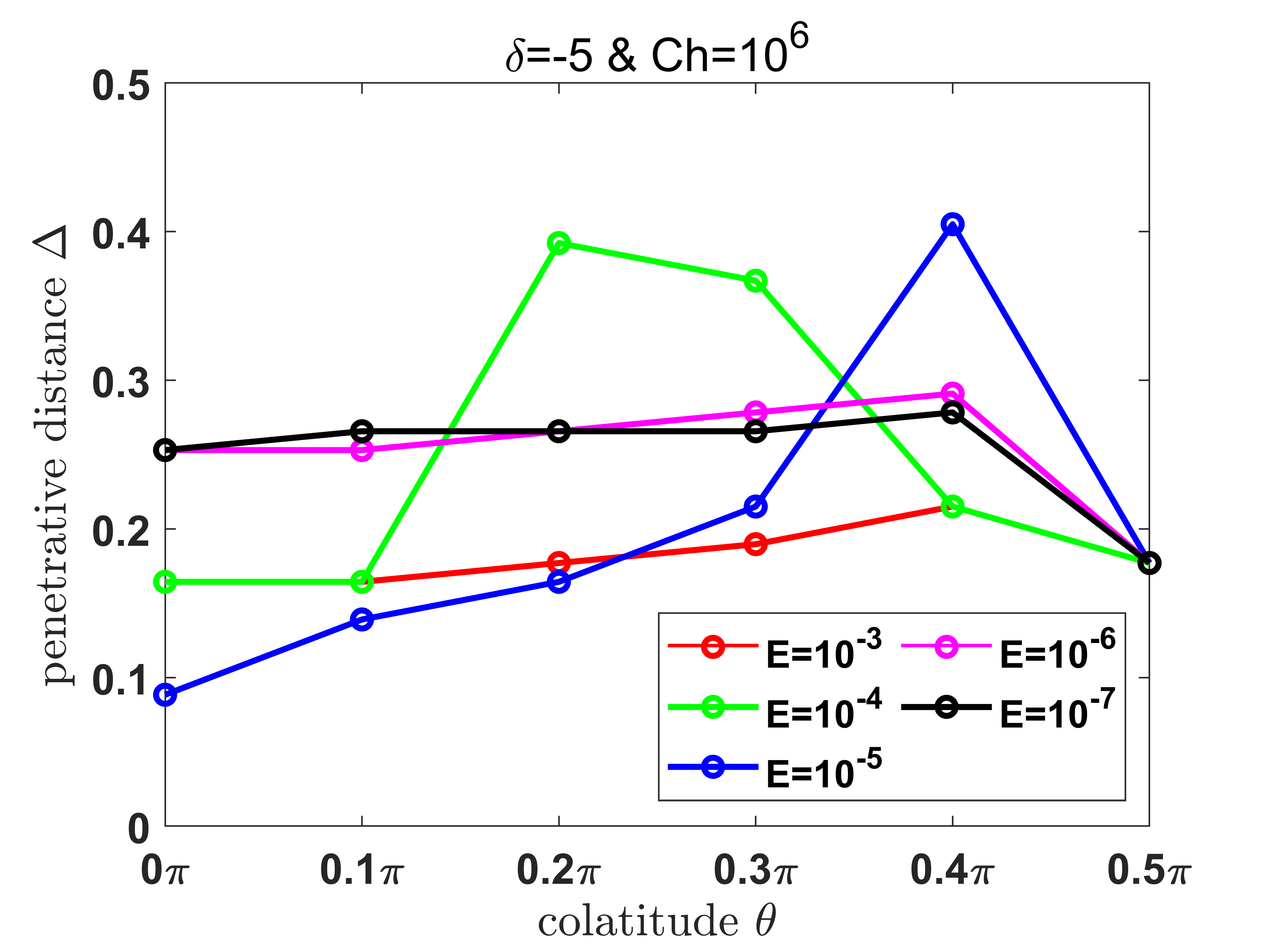}}
    \subfigure[]{
    \includegraphics[width=0.45\columnwidth]{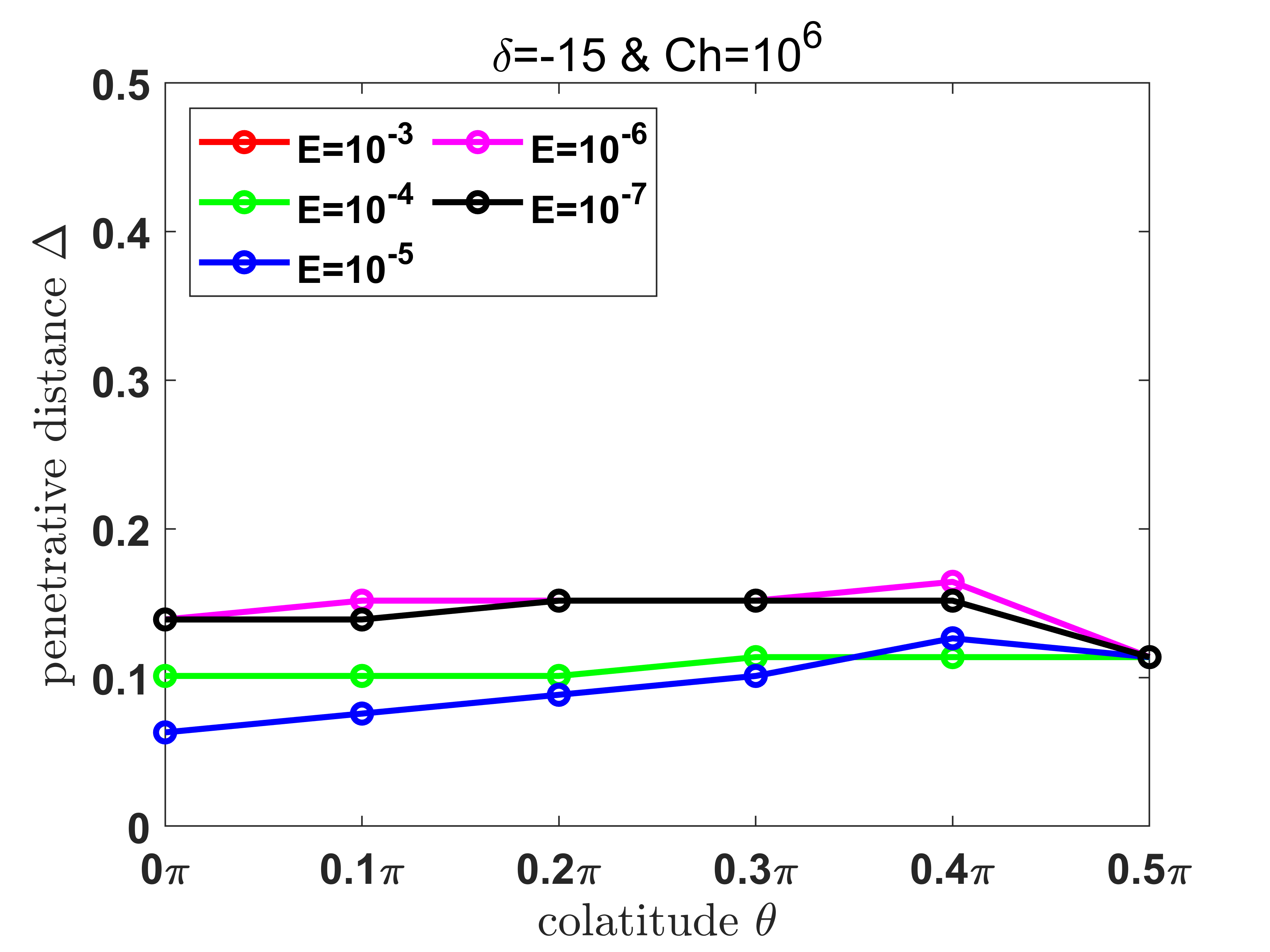}}
    \caption{Companion to Fig.~\ref{fig:f3} but with a perfect conduct boundary condition at the bottom.}
    \label{fig:f9}
\end{figure}

\begin{figure}
    \centering
    \subfigure[]{
    \includegraphics[width=0.45\columnwidth]{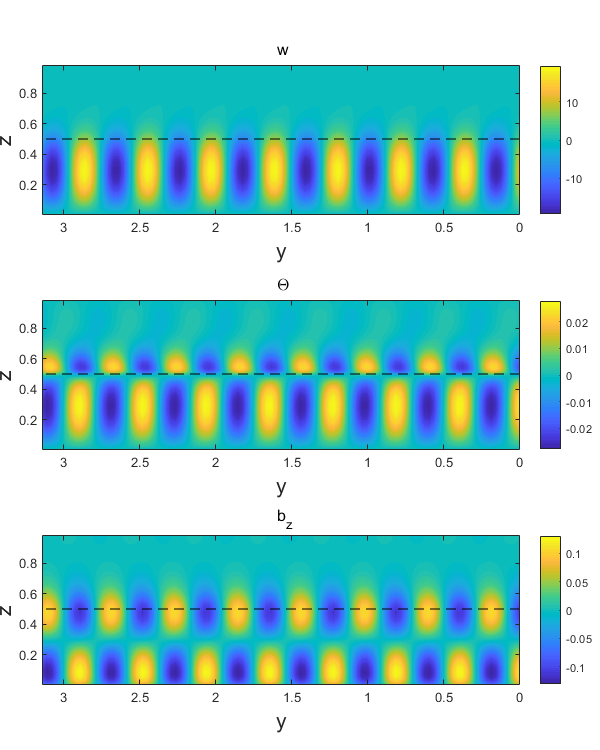}}
    \subfigure[]{
    \includegraphics[width=0.45\columnwidth]{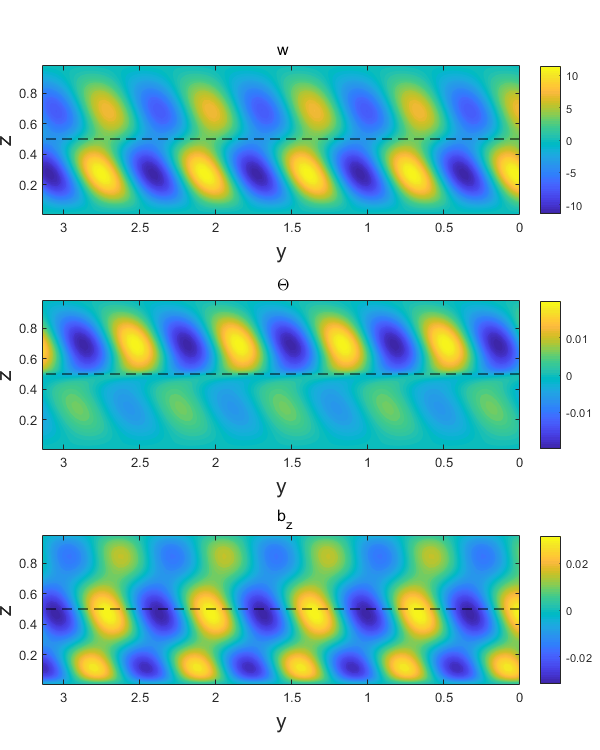}}
    \subfigure[]{
    \includegraphics[width=0.45\columnwidth]{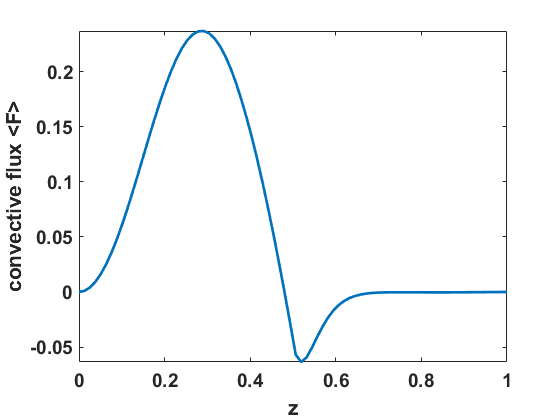}}
    \subfigure[]{
    \includegraphics[width=0.45\columnwidth]{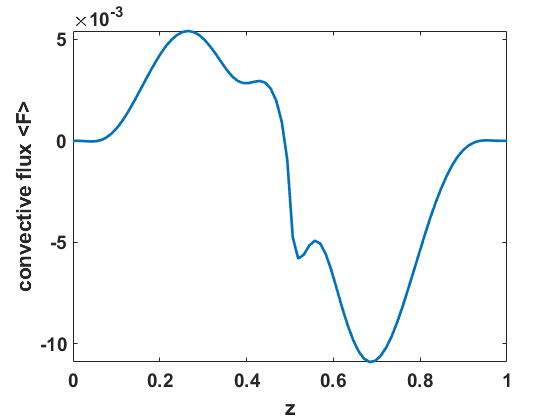}}
    \caption{The contour plots for the vertical velocity $w$, the temperature perturbation $\Theta$, and the vertical component of magnetic perturbation $b_{z}$, at $\delta=-5$, $\theta=0.2\pi$, $Pr=Pm=0.1$, $E=10^{-4}$, $Ch=10^{6}$ and $\gamma=\theta$. The left and right columns use different magnetic boundary conditions at the bottom with (a)psuedo-vacuum and (b)perfectly conducting, respectively. (c) and (d)The corresponding convective flux $\langle F\rangle$ as a function of $z$.}
    \label{fig:f10}
\end{figure}

Fig.~\ref{fig:f9} presents results for cases employing a perfectly conducting boundary condition at the bottom. When compared to Fig.~\ref{fig:f3}, it becomes evident that the penetrative distances are marginally shorter under the perfectly conducting boundary condition than under the pseudo-vacuum boundary condition, especially for weak background magnetic field. Some unusual cases arise at low to mid-latitudes when $\Lambda > 1$. For instance, with the use of a perfectly conducting boundary condition, the penetrative distance undergoes a sudden shift at $\theta=0.2\pi$ for the group $E=1\times 10^{-4}$, $Ch=1\times 10^{6}$, and $\delta=-5$. To elucidate this, we have depicted the results for both pseudo-vacuum and perfectly conducting boundary conditions in the left and right columns of Fig.~\ref{fig:f10}, respectively. In the case of the perfectly conducting condition, wave-like structures become apparent for $w$, $\Theta$, and $b_{z}$. Interestingly, $\Theta$ exhibits greater strength in the stable layer than in the unstable layer, leading to a flux magnitude in the stable layer that can be twice as large as that in the unstable layer. As demonstrated in these figures, the penetration process can become complex when $\Lambda \gg 1$, due to the emergence of additional magnetic waves. As a result, the change of magnetic boundary condition can potentially lead to substantial modifications in the structure, and the penetrative distance as well.

%
%

%


\bibliography{paper}

\end{document}